\documentclass[superscriptaddress,nofootinbib,a4paper,11pt]{article}
\usepackage{jheppub}
\usepackage{graphicx}
\usepackage{subcaption}
\usepackage{float}
\usepackage{amsmath}
\usepackage{amssymb}
\usepackage[utf8]{inputenc}
\usepackage{hyperref}
\usepackage{color}

\begin{document}

\title{\boldmath Reconstructing boosted Higgs jets from event image segmentation}

\author[a]{Jinmian Li,}
\author[b,c]{Tianjun Li,}
\author[d,b]{and Fang-Zhou Xu}

\affiliation[a]{College of Physics, Sichuan University, Chengdu 610065, China}
\affiliation[b]{CAS Key Laboratory of Theoretical Physics, Institute of Theoretical Physics, Chinese Academy of Sciences, Beijing 100190, China}
\affiliation[c]{School of Physical Sciences, University of Chinese Academy of Sciences,
No.~19A Yuquan Road, Beijing 100049, China}
\affiliation[d]{Center for High Energy Physics, Tsinghua University, Beijing 100084, China}

\emailAdd{jmli@scu.edu.cn}
\emailAdd{tli@mail.itp.ac.cn}
\emailAdd{xfz14@mails.tsinghua.edu.cn}

\abstract{
Based on the jet image approach, which treats the energy deposition in each calorimeter cell as the pixel intensity, the Convolutional neural network (CNN) method has been found to achieve a sizable improvement in jet tagging compared to the traditional jet substructure analysis. 
In this work, the Mask R-CNN framework is adopted to reconstruct Higgs jets in collider-like events, with the effects of pileup contamination taken into account. This automatic jet reconstruction method achieves higher efficiency of Higgs jet detection and higher accuracy of Higgs boson four-momentum reconstruction than traditional jet clustering and jet substructure tagging methods. 
Moreover, the Mask R-CNN trained on events containing a single Higgs jet is capable of detecting one or more Higgs jets in events of several different processes, without apparent degradation in reconstruction efficiency and accuracy. 
The outputs of the network also serve as new handles for the $t\bar{t}$ background suppression, complementing to traditional jet substructure variables. 


 }

\maketitle

\section{Introduction}

Jet is one of the most prominent objects in the event reconstruction at the Large Hadron Collider (LHC). 
Due to the color confinement, quark/gluon can not be detected freely. 
Almost immediately after being produced, a quark/gluon goes through parton showering and hadronization, leading to a collimated spray of energetic detectable particles, which is referred to as jet~\cite{PhysRevLett.35.1609, Blazey:2000qt, Ellis:2007ib, Salam:2009jx}. 
At a high energy collider such as the LHC, a boosted hadronically decaying heavy particle can also give rise to a jet, e.g., W/Z boson, Higgs boson and top quark in the Standard Model (SM).  

The aims of jet reconstruction are to obtain the original parton~\footnote{The parton here also refers to a boosted hadronically decaying W/Z boson or  Higgs boson.} momentum and identity by using the information of final state hadrons. 
Since the first proposal for jet reconstruction~\cite{Sterman:1977wj}, many jet clustering algorithms have been developed and adopted in experiments. 
At lepton colliders and hadron colliders, jets are usually reconstructed through a sequential recombination algorithm, such as the $k_T$ algorithm~\cite{Ellis:1993tq,Catani:1993hr}, the Cambridge/Aachen algorithm~\cite{Dokshitzer:1997in}, the anti-$k_t$ algorithm~\cite{Cacciari:2008gp}. 
Those algorithms involve a cone size parameter ($R$) that should be adjusted according to the detector architecture and the property of the target jet. 
The primary task after jet clustering is to identify the jet origin (jet tagging), for which the jet substructure is found to be very useful. Many dedicated non-machine-learning variables and methods (see Refs.~\cite{Abdesselam:2010pt, Altheimer:2012mn, Altheimer:2013yza,Adams:2015hiv, Asquith:2018igt} for reviews) have been proposed to distinguish top quark jet from light flavor jet~\cite{Kaplan:2008ie, Plehn:2009rk, Plehn:2010st, Soper:2012pb}, W/Z/H jet from QCD jet~\cite{Butterworth:2008iy}, as well as quark jet from gluon jet~\cite{Gallicchio:2011xq, Gallicchio:2012ez}. 
Despite the great success of the jet clustering algorithm, there are several possible issues remain: (1) The jet clustering can be easily infected by another closeby hard parton which could come from multiple minimum-bias interactions (pileup) or from underlying events~\footnote{The jet grooming techniques~\cite{Butterworth:2008iy,Krohn:2009th, Ellis:2009su} can mitigate those effects.}; (2) The cone-size is an a-priori parameter in jet clustering. Hadrons that have different origins may be clustered in the same jet if the $R$ is too large, or those have the same origin may be mis-clustered into different jets  if the $R$ is too small. In both cases, the jet substructure is distorted and the jet tagging performance is reduced. 

The recent developments of computer vision, especially the application of deep learning~\cite{Larkoski:2017jix,Guest:2018yhq, Albertsson:2018maf, Radovic:2018dip} (see Ref.~\cite{Feickert:2021ajf} for more references), can be used to reconstruct and tag the jet nature with low-level inputs (four-momentum vectors of final state hadrons). 
From physics of jet formation, jets can be viewed as sequences~\cite{Andreassen:2018apy} and trees~\cite{Louppe:2017ipp,Cheng:2017rdo} formed through sequential parton showering, or viewed as graphs~\cite{Moreno:2019bmu,Guo:2020vvt,Dreyer:2020brq} and point cloud~\cite{Komiske:2018cqr,Qu:2019gqs,Dolan:2020qkr} with the information of jet constituents encoded in the adjacency matrix and node features. 
These jet representations are found to be successful in tagging the jet identity with deep neural network (DNN). 
Inside a detector, the calorimeters measure the positions and energy depositions of jet constituents on fine-grained spatial cells. 
Treating each cell as a pixel, and the energy deposit in the cell as the intensity (or grayscale color) of that pixel, the jet can be naturally viewed as a digital image. 
There are many works that use the image-based approach for various jet tagging tasks, e.g., W/Z jet tagging~\cite{Cogan:2014oua,deOliveira:2015xxd,Baldi:2016fql}, top quark jet tagging~\cite{Almeida:2015jua, Kasieczka:2017nvn,Macaluso:2018tck, Kasieczka:2019dbj,CMS-PAS-JME-18-002}, Higgs jet tagging~\cite{Lin:2018cin,Chung:2020ysf}, quark-gluon jet discrimination~\cite{Komiske:2016rsd,ATL-PHYS-PUB-2017-017} and new heavy particle jet tagging~\cite{Guo:2018hbv}. 
In these methods, the traditional jet clustering algorithm is used to reconstruct the jet in an event and a DNN is applied subsequently for jet identification. 
Apparently those methods suffer from similar issues as in jet clustering, i.e., contaminations from soft-radiation and pileup events, jet image distortion due to inappropriate cone-size parameter. 
Moreover, in our previous study~\cite{Guo:2018hbv}, we find the neutralino ($\tilde{\chi}$) jet tagging efficiency of the neural network which is trained on the event sample of $\tilde{\chi} \tilde{\chi}+$jets process degrades when applying to events of other processes. 
This implies that the network learns some process dependent features of the jet and the problem is more severe as the final state multiplicity is higher. 
The performance of the network being as process agnostic as possible is an important issue that should be addressed during the training. 

A natural extension of the DNN application to jet tagging is implementing the jet detection within the DNN, so that the manual jet clustering is not needed. 
The techniques of object detection and image segmentation in computer vision is one of the possible solution. There are mainly two types of object detection methods: (1) the region proposal based framework such as Mask R-CNN~\cite{mask}; (2) the regression/classification based framework such as Yolo~\cite{2015arXiv150602640R, 2018arXiv180402767R}. 
In this work, we adopt the Mask R-CNN architecture to detect (reconstruct) the Higgs jet in Higgs events which is overlaid with dozens of pileup events. The loss function of the network is designed to achieve the highest precision of the Higgs jet reconstruction. 
We find this automatic Higgs detection method outperforms the traditional jet clustering and jet substructure tagging method, in the sense that the Higgs jet detection efficiency, the Higgs jet 4-momentum reconstruction precision, and the background rejection rate are higher. 
Moreover, we show that the Mask R-CNN trained on single Higgs events is capable of detecting one or more Higgs jets in different processes. 

The paper is organized as follows. In Sec.~\ref{sec:data} and Sec.~\ref{sec:network}, we introduce the data preparation and the architecture of the network. The performances of the CNN method are presented in Sec.~\ref{perform}. We provide conclusion and outlook in Sec.~\ref{sec:conclusion}. 

\section{Event generation and data preparation}  \label{sec:data}

Event samples of the $H$+jets process with flat Higgs transverse momentum ($p_T^H$) distribution are generated by MG5\_aMC@NLO~\cite{Alwall:2014hca} for training and validating our network. The Higgs boson is forced to decay into a pair of $b$-quark. 
We consider the $p_T^H$ in the range of [200,600] GeV, so that the Higgs boson forms a resolvable jet. In each bin with width of 25 GeV, 50K events are generated to guarantee the flatness of the $p_T^H$ distribution. 
Note that $p_T^H$ requirements are applied at the parton level.
The Higgs momenta are altered a little by initial-state showers, whose simulations are handled by Pythia8~\cite{Sjostrand:2007gs}.
Then 50 pileup events~\footnote{We use the fixed number of pileup events as a conservative estimation. At the last stage of the LHC run 2, the average pileup event number is $\sim 35$.} are superposed on each one of the hard events. 
Our network is trained with 640K events and validated on 160K events. 
The test samples in the following discussion are generated separately with 100K events in each Higgs $p_T$ bin.



\begin{figure}[htbp]
\centering
\begin{subfigure}[t]{0.48\textwidth}
\includegraphics[width=\textwidth]{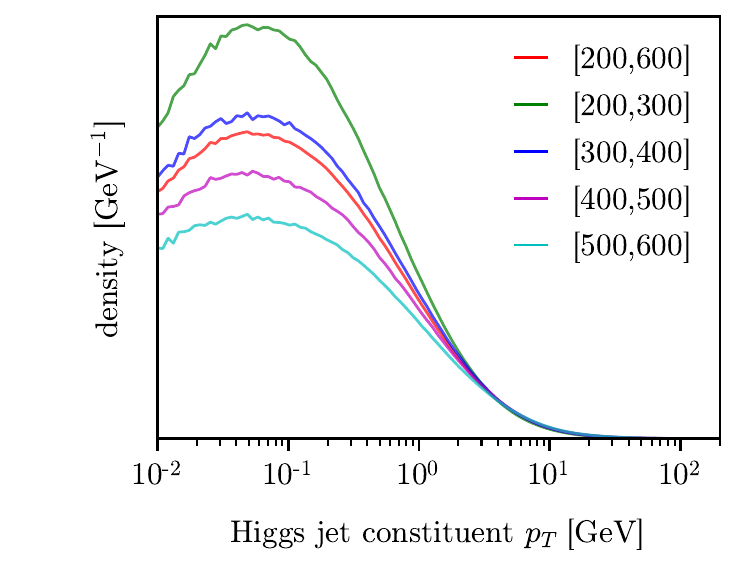}
\end{subfigure}
\begin{subfigure}[t]{0.48\textwidth}
\includegraphics[width=\textwidth]{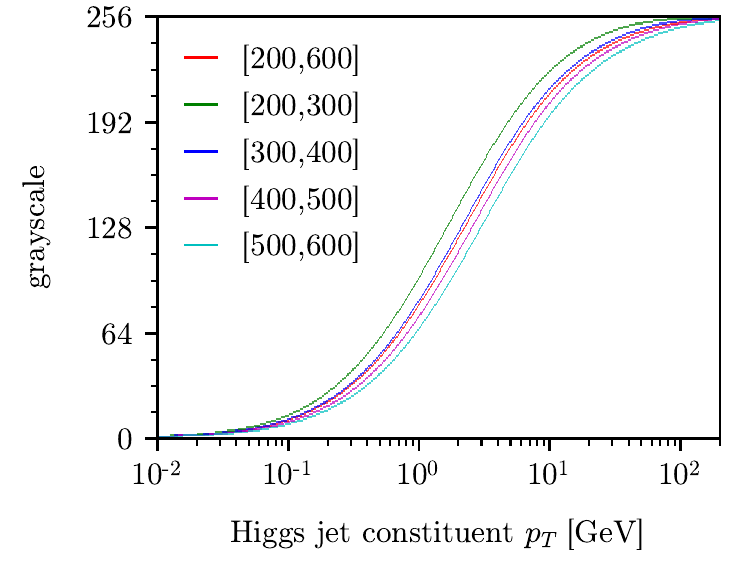}
\end{subfigure}
\caption{Left panel:  transverse momentum distributions of Higgs decay products for training samples in different $p_T^H$ ranges (in unit of GeV). Right panel: pixel intensity normalization schemes for different $p_T^H$ samples.}
\label{fig:pt_higgs}
\end{figure}

The data fed to the network are in the form of 320$\times$320 grayscale images, representing the transverse momentum deposited in the $\eta \times \phi$ plane across $[-\pi,~\pi]\times[0,2\pi]$. Each pixel corresponds to $\Delta\eta\times\Delta\phi\simeq0.020\times0.020$. 
Since the image grayscale are integers that range from 0 (black) to 255 (white), we need to set the correspondence between the greyscale and particle $p_T$. 
Given a Higgs jet, the momenta of its constituents span over 4 orders of magnitude, concentrated at low $p_T$ region. In order to resolve the particles with low $p_T$, one simple way to represent them is a logarithmic transformation. But the consequence is still lack of sensitivity to soft radiation and waste of intensity gradient in regions above 10 GeV, i.e. too many particles in lower $p_T$ bins and too few particles in higher $p_T$ bins. 
In order for the images to exploit as many $p_T$ scales as possible, the 256 bins of $p_T$ (with each bin corresponds to one grayscale value) are chosen such that number of final state particles (the Higgs jet constituents) in each bins are approximately the same. 
The procedure can be thought of as an image enhancement method, analogous to processing high dynamic range (HDR) images with the tone mapping technique to increase local contrast~\cite{enhance}. 
The Fig.~\ref{fig:pt_higgs} shows the $p_T$ distributions of the Higgs decay products and the grayscale mappings for different $p_T^H$ samples (of the $H+$jets process). 
Numerically, curves on the right panel equal the ones on the left panel being integrated and multiplied by 256, which is just 8 bit color depth. 
An image represented in this way can maximize the $p_T$ information that can be encoded by greyscale. 
Note that there are only slight differences between normalizations of different $p^H_T$ samples, as shown in the right panel of Fig.~\ref{fig:pt_higgs}.
So a unified normalization scheme is acceptable.
In our study, the normalization which is obtained from the event sample with $p^H_T \in [200~\text{GeV},600~\text{GeV}]$ is used. 
Moreover, although this procedure is optimized based on Higgs jets of certain energy, it should be applicable to other objects to some extent, e.g., hadronically decaying vector bosons and top quarks, since $p_T$ distributions of jet constituents depend largely on the energy scales of parton shower and hadronization rather than identities of the particles which initiate the jets.

\begin{figure}[htbp]
\centering
\includegraphics[width=0.24\textwidth]{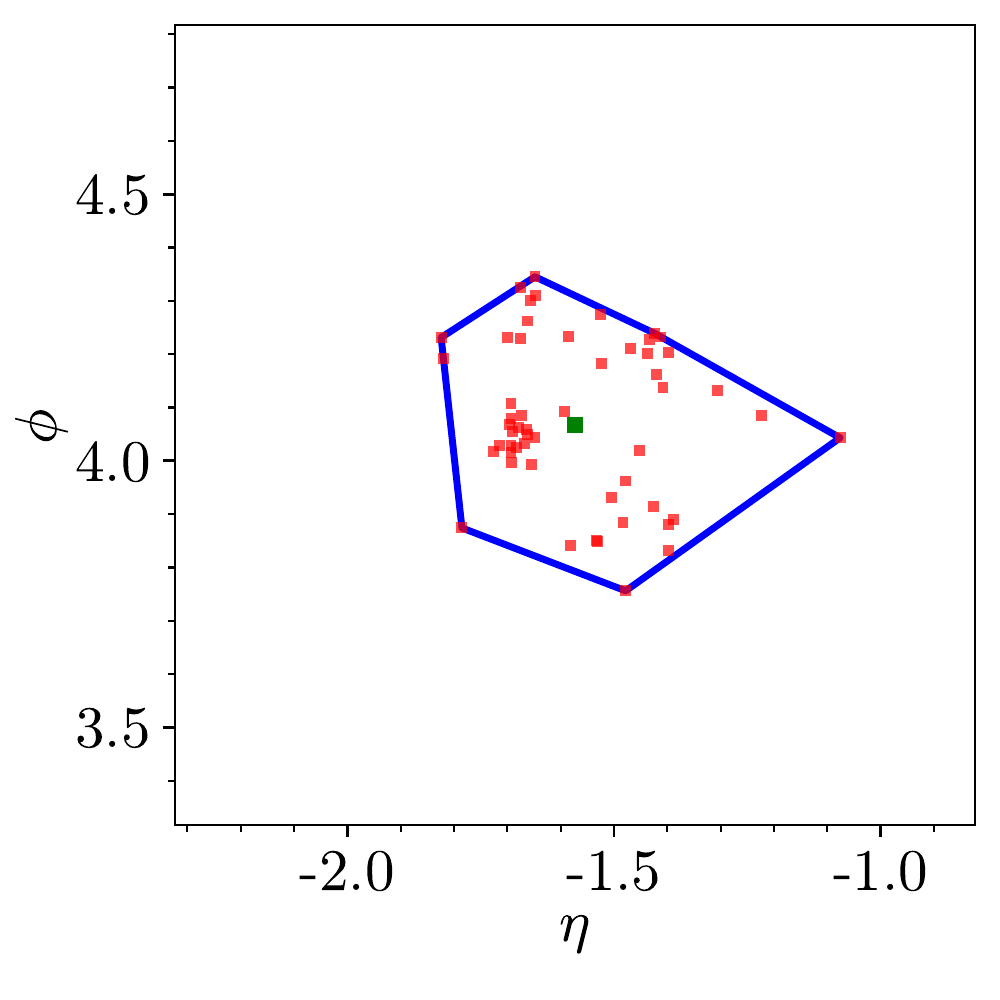}
\includegraphics[width=0.24\textwidth]{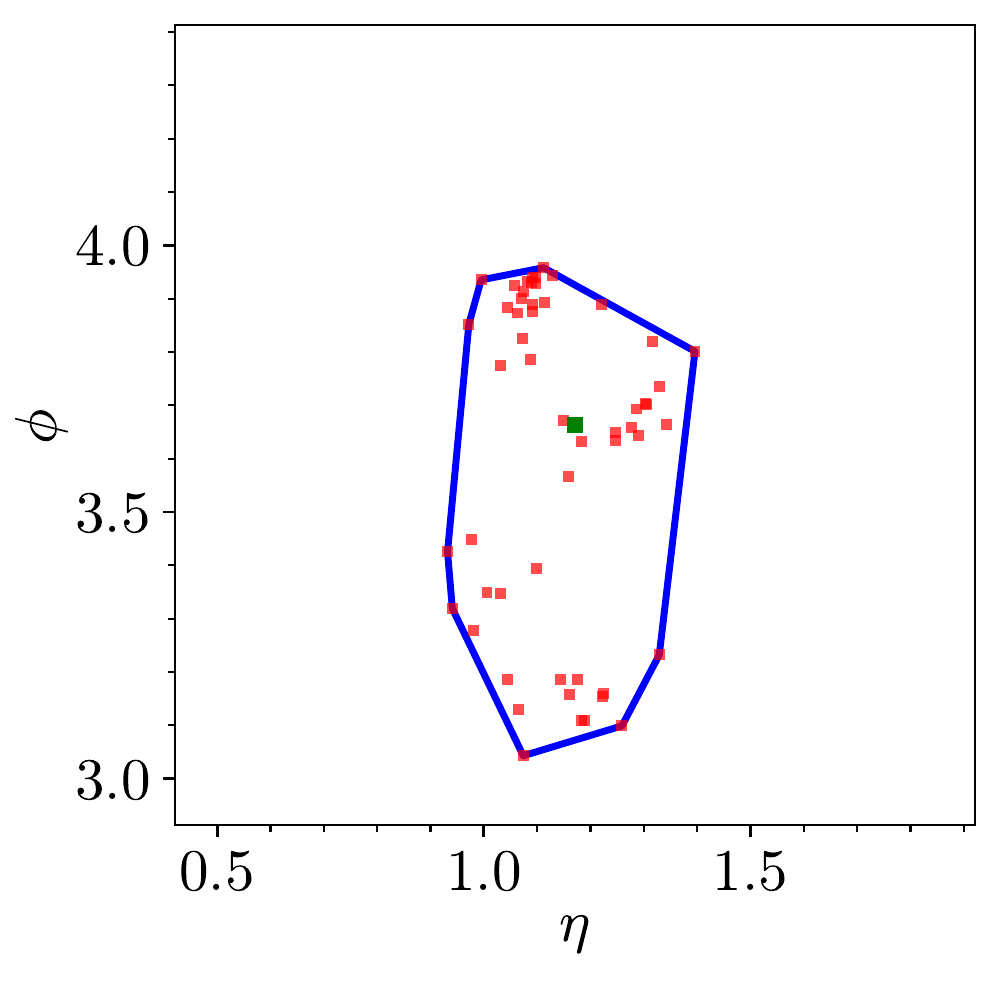}
\includegraphics[width=0.24\textwidth]{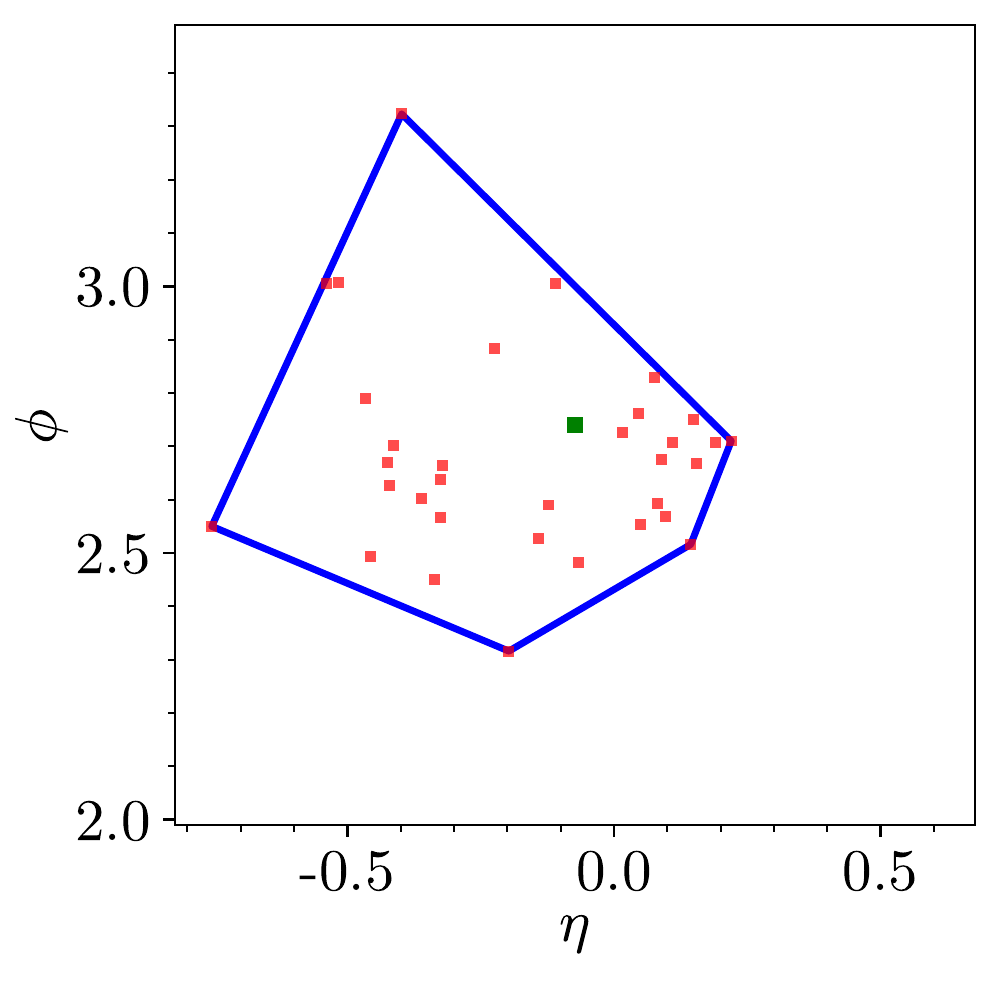}
\includegraphics[width=0.24\textwidth]{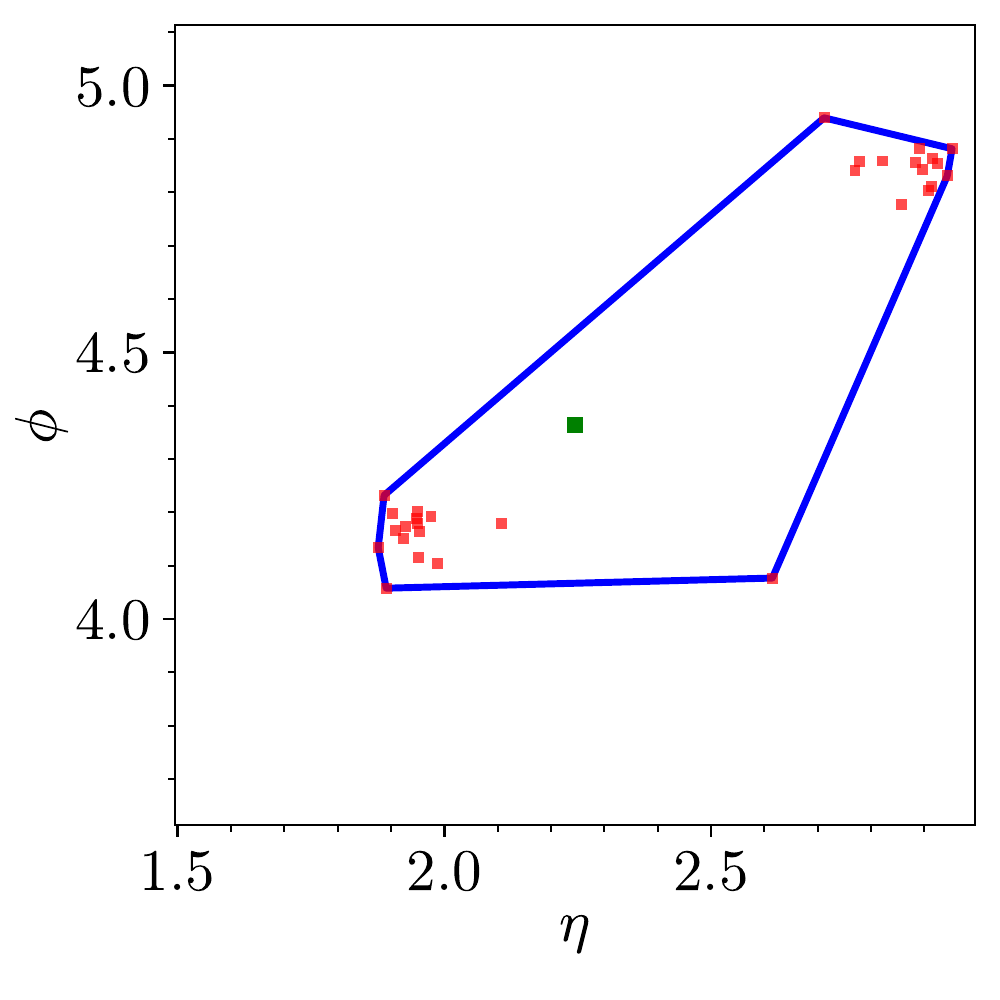}
\includegraphics[width=0.24\textwidth]{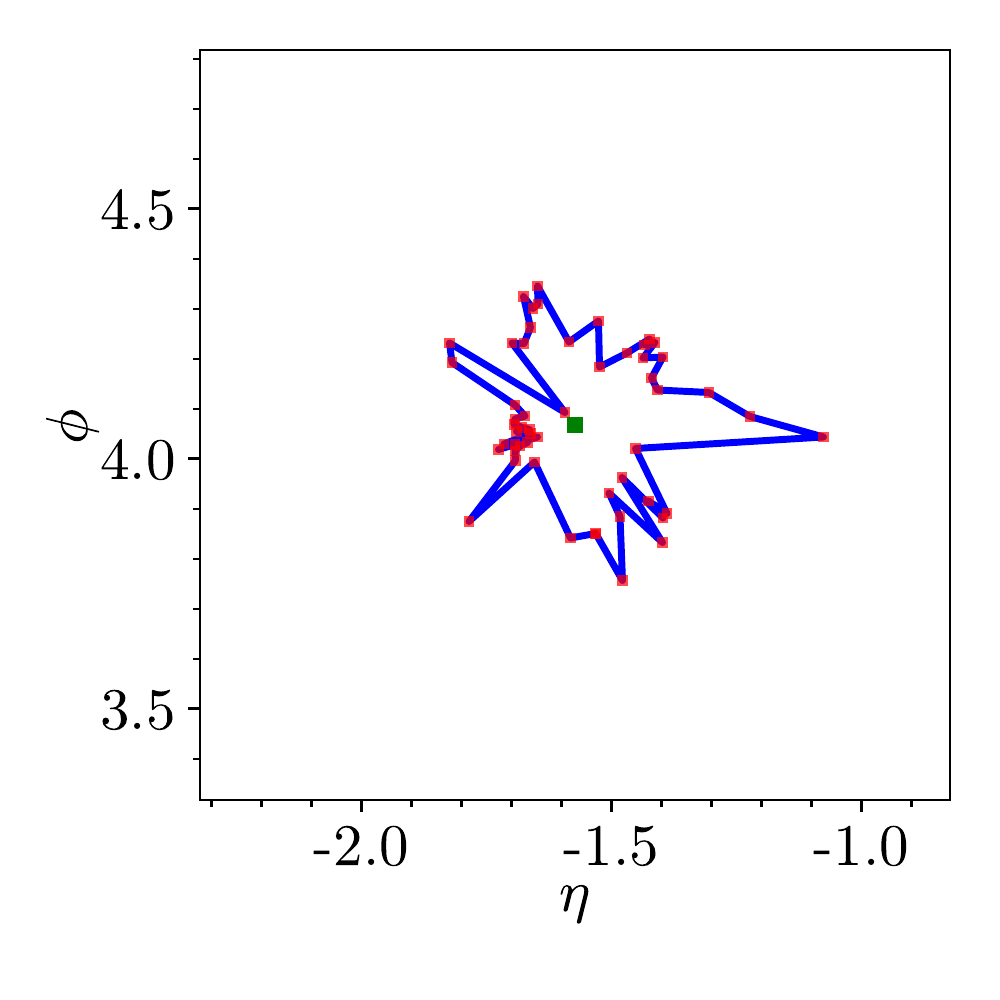}
\includegraphics[width=0.24\textwidth]{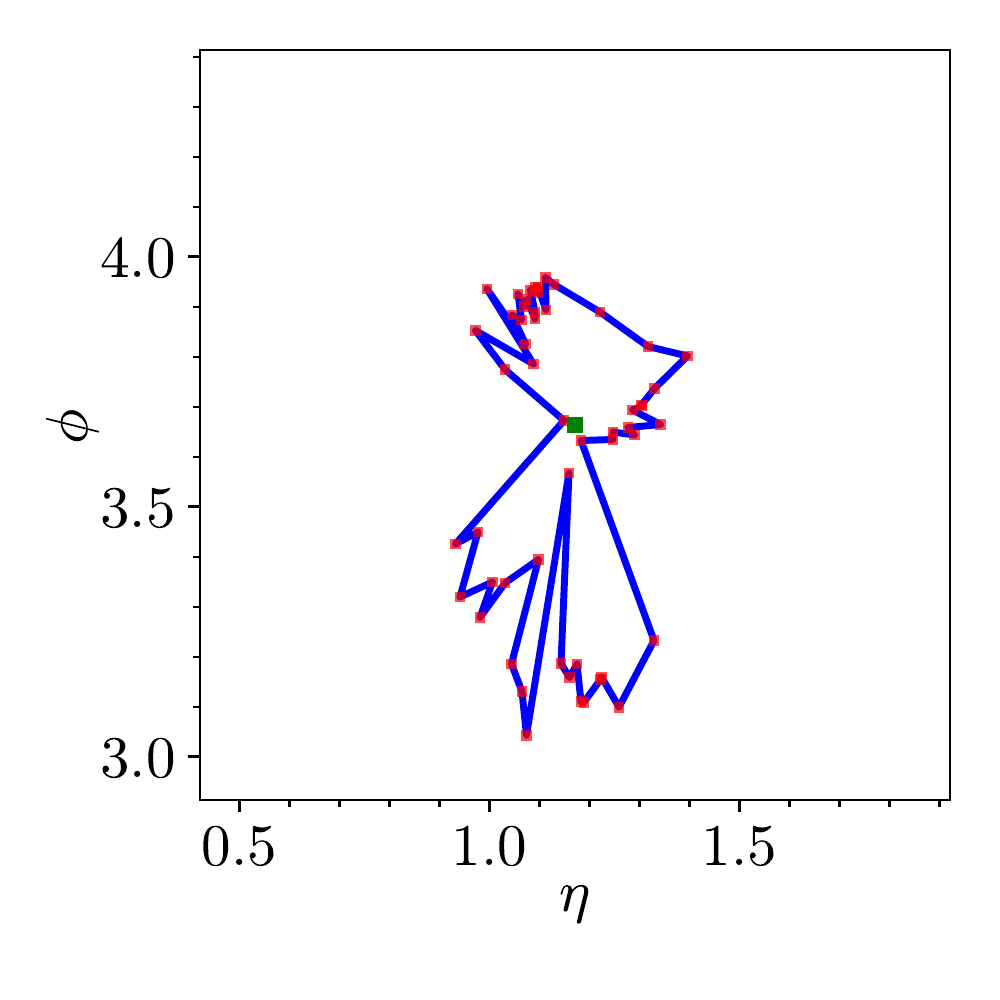}
\includegraphics[width=0.24\textwidth]{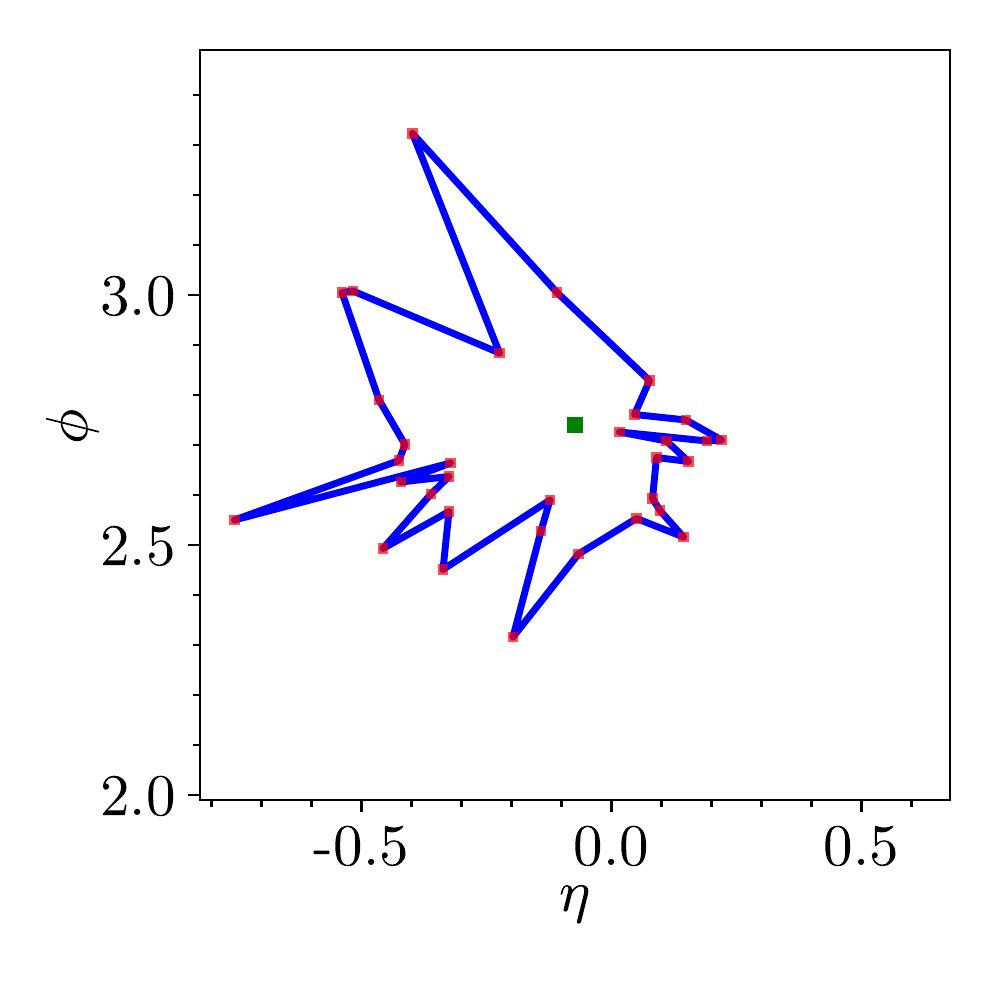}
\includegraphics[width=0.24\textwidth]{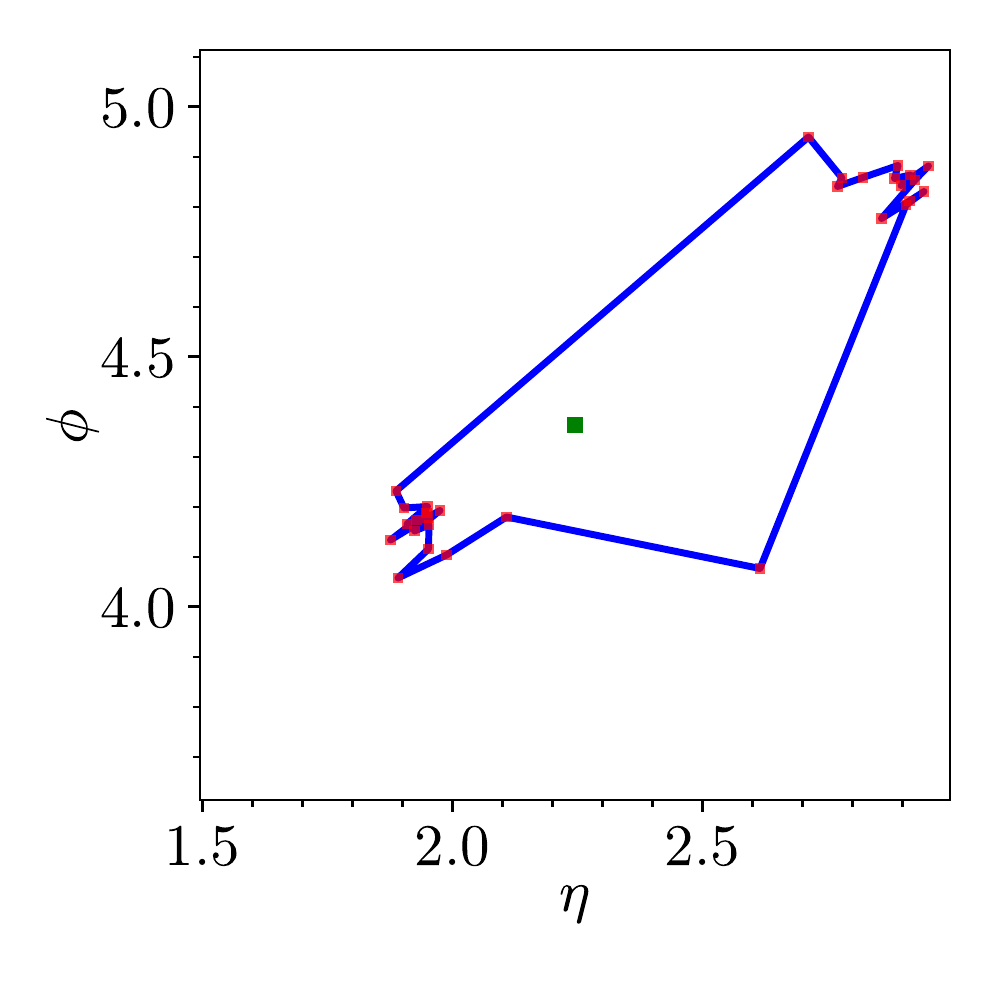}
\caption{Top panels: convex hull masks. Bottom panels: radial masks. Green dots represent $y\times \phi$ coordinates of Higgs bosons, while red dots represent $\eta \times \phi$ coordinates of their decay products.  }
\label{fig:masks}
\end{figure}

To tell the network the location of the Higgs jet on a detector image, each image is complemented by a mask in the form of 320$\times$320 binary image (in the case of multiple Higgs jets, several masks) where signal pixels are labeled 1 and the rest 0. 
Technically, only individual pixels representing particles coming from the Higgs boson are supposed to light up, no matter their $p_T$, which is obviously easy to implement. The issue is that our network, heavily relying on convolution, could not handle masks composed of sparse pixels very well. On the contrary, a connected mask is preferred, i.e., a jet area. Though this kind of labeling could render the algorithms more susceptible to pile-up, we will see that, due to its ability to keep the jet area noticeably smaller than conventional jet clustering algorithms, the influences are not significant and can be estimated. 
An appropriate pileup mitigation procedure based on jet areas will make the four-momenta of reconstructed jets more accurate. We present two schemes of defining masks.
They share common pre-selection rules. Boost along the beam direction to the frame where $p_z$ of the Higgs boson becomes zero, then discard all constituent particles with energy lower that 1 GeV or with angular separation to the Higgs boson ($\Delta \theta$) greater than $\pi/2$. The Higgs jets after those two selections are referred to as the ground truth (GT) Higgs.
A simple alternative is requiring $p_T>1$ GeV and $\Delta R=\sqrt{\Delta \phi^2 + \Delta \eta^2}<\pi/2$ (instead of energy and angular separation), but it does not treat particles with different orientations on an equal footing. The inequalities are more manifest when decay products are scattered more widely on the $\eta \times \phi$ plane due to higher rapidities or lower $p_T$ of Higgs bosons.
Note that only differences in rapidity $\Delta y$ are invariant under boosts along the beam direction, thus the proper measure of relative locations of particles on an image should be $y\times \phi$ instead of $\eta \times \phi$. Even though $y$ can be approximated by $\eta$ for ultrarelativistic particles, this approximation does not hold for Higgs boson. 
Therefore the coordinates representing a Higgs boson on the $\eta \times \phi$ plane are taken to be its rapidity $y$ and azimuth $\phi$. 
The first mask is a regions bounded by the convex hull covering all the constituents, while the second mask is a region bounded by an irregular polygon with serrated edge. We construct it as follows; sort the constituents according to their polar angle with respect to the initial Higgs boson, then connect them sequentially in a closed loop, to form the countour of the radial mask. The surface of this mask is bound to be equal to or smaller than that of the convex hull in size. 
The masks defined in those ways are simply connected regions, so that a convolutional network can be applied efficiently. 
Moreover, the area of a mask is typically much smaller than that of a jet reconstructed from jet clustering algorithm. A quantitive result is shown in Sec.~\ref{sec:recopile}. This feature helps to reduce the pileup contamination. 
During the testing, a successful Higgs jet detection means that the $y \times \phi$ coordinates of the ground truth Higgs boson lies within the mask. 
Other ways to construct a mask are possible and could potentially boost the performance. We show some instances in Fig. \ref{fig:masks}. Note that the pixelation of edges due to finite spacial resolution is not considered at this stage. 

Considering that some jet constituents (in the case of radial masks, all jet constituents) are located right on edges of masks and that it is harder for a CNN to be accurate down to pixel level, we enlarge the predicted convex hull (radial) mask by one (two) pixel(s) around its boundary during test. 
Enlarging by one pixel is implemented by adding the neighbors ($P_{(i-1)j}$, $P_{i(j-1)}$, $P_{(i+1)j}$ and $P_{i(j+1)}$ which do not belong to the mask) of each masked pixel ($P_{ij}$) into the mask.
The same procedure can be applied recursively to enlarge the mask by two pixels. 
The size of enlargement is determined by whether the distributions of the transverse momentum ($p_T$) and the invariant mass ($m$) of the masked jet are changed significantly. We find the changes of $p_T$ and $m$ distributions are noticeable when the enlargement is one (two) pixel(s) for convex hull (radial) mask, while the changes are much milder for further enlargement. 

\section{Network Architecture} \label{sec:network}

\begin{figure}[t!]
\centering
\includegraphics[width=\textwidth]{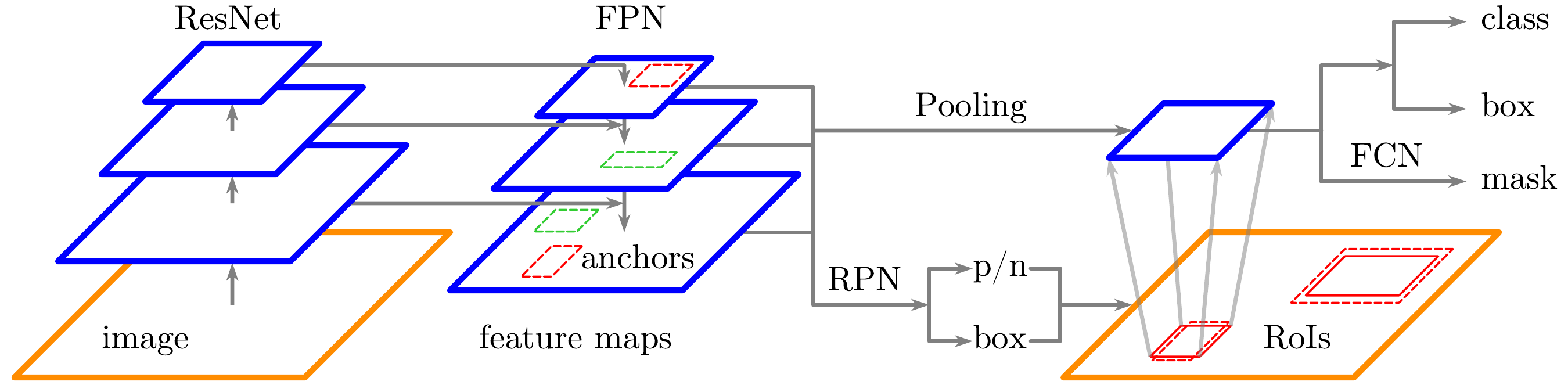}
\caption{A sketch illustrating the components of Mask R-CNN and its processing pipeline, inspired by figures in Refs. \cite{rcnn,fast,faster,mask}. \label{fig:maskrcnn} }
\end{figure}


CNNs are well suited for image recognition tasks~\cite{10.1145/3065386,DBLP:journals/corr/SimonyanZ14a,7780459}. Basically, convolution layers compute feature maps at different levels, pooling layers perform downsampling on them, and fully-connected layers are used for regression or classification. These are basic building blocks of a CNN. Different network architectures can be constructed by stacking multiple layers in various combinations to accomplish all kinds of recognition tasks, such as classification, detection and segmentation. Certain networks also incorporate deconvolution layers for upsampling, so that the spatial dimension of data can be increased.
Convolution operations work in a translation-invariant manner. The nature of images and CNNs correspond exactly to the finite space resolution of a detector and the Lorentz boost invariance of $y-\phi$ coordinates, making them well suited for the jet detection.

Mask R-CNN~\cite{mask} is a framework extensively adopted in the computer vision industry for object detection and semantic segmentation. It was developed progressively from the region-based CNN (R-CNN) framework~\cite{rcnn} and first put forth in 2017.
Mask R-CNN is created by intricately combining three major functional modules, a region proposal network (RPN), a Region-based CNN (R-CNN) and a fully convolutional network (FCN)~\cite{7298965}. Both RPN and R-CNN are adapted by replacing the single-scale feature map with a feature pyramid network (FPN)~\cite{8099589} to detect multi-scale objects. In the original paper, different convolutional architectures in the R-CNN for feature extraction are compared. We choose the residual neural network~\cite{7780459} of 50 layers (ResNet-50) as the convolutional architecture. We also test a deeper network, ResNet-101, and observe no significant improvements.

The workflow for the Mask R-CNN is illustrated in Fig.~\ref{fig:maskrcnn}. 
The orange square represents an input image and blue squares represent feature maps of multiple scales. 
In the R-CNN module, the ResNet-50 is used for feature extraction. The sketch shows only three levels instead of five in the actual implementation. The early level detects the local features, and later level successively detects the global features. 
Each map in FPN is upsampled and merged with the corresponding map in ResNet to generate a lower-level one, such that every level has the access to both local and global features. 
Anchors (dashed rectangles in the sketch) are a set of reference boxes with fixed scales and aspect ratios that spread across the feature pyramid. 192 anchors with scale of 64 pixel and aspect ratio of 1:1, 1:2 and 2:1 (64 anchors for each ratio) are randomly generated in each feature map of FPN. 
The design of FPN and anchors enable Mask R-CNN to deal with multi-scale objets. This feature is crucial for tagging Higgs jets with varying transverse momenta.
The RPN consists of two sibling branches: a classification branch that aims at predicting the label of anchors defined according to its Intersection over Union (IoU) with ground-truth box (positive for $\text{IoU}>0.5$ and negative otherwise, shown by red and green dashed rectangles in the Fig.~\ref{fig:maskrcnn}), and a regression branch that refines the boxes so as to match the target boxes defiened with maximum IoU. 
The RPN outputs a set of rectangular proposals, referred to as regions of interest (RoIs, solid red rectangles in the sketch), exploiting the hierarchical features computed by the R-CNN.
The RoI pooling layer converts the section of feature map corresponding to each (variable sized) RoI into fixed size~\footnote{This is implemented with {\it tf.crop\_and\_resize()} operation in tensorflow, using bi-linear interpolation~\cite{DBLP:journals/corr/JaderbergSZK15}.}.
A fully-connected layers perform classification and bounding-box regression on the pooled feature map (as in RPN, the target is to obtain the smallest bounding box for the Higgs jet, {\it i.e.} $\text{IoU} \to 1$). 
In our case, there are only two classes, Higgs jet and background. The features are shared between the proposal and detection networks. 
In parallel to predicting the class and bounding-box, a small FCN outputs a binary mask determining whether a pixel belongs to the jet.
Though seemingly simple in concept, Mask R-CNN is quite a sophisticated framework to implement. Fortunately, source codes of at least three different implementations have been made publicly available. We use the one at \url{https://github.com/matterport/Mask_RCNN}. For elaboration of concepts and implementation details, see Ref. \cite{mask} and references therein.

A detector image has slightly different topology than an ordinary one, i.e., $\phi=0$ and $\phi=2\pi$ represent the same line. The decay products of a Higgs boson with $\phi\gtrsim 0$ or $\phi\lesssim 2\pi$ will locate in two separate regions on an image, making the jet difficult to detect and cluster. 
The Mask R-CNN method can not be applied to detect such objects directly. 
In order to accommodate the cylindrical topology in a CNN, one need to incorporate a periodic padding feature into the underlying framework (in our case, TensorFlow), which is however complicated and beyond the scope of current work. 
We opt for another approach to bypass this challenge. If the $\phi$-coordinate of a Higgs boson lies outside of $[\pi/2,3\pi/2]$, then a shift of $\pm \pi$ is applied to $\phi$ accordingly to move the Higgs boson into $\phi\in[\pi/2,3\pi/2]$. Combined with the pre-selection rule that the angles between a Higgs boson and its jet constituents be smaller than $\pi/2$, this procedure ensures the unity of a mask. This trick works for cases of at most two Higgs bosons in one event.
Note that without such $\phi$-warping approach, our CNN will not be able to detect Higgs jets across the boundary ($\phi=0$). 

\section{Network Performance} \label{perform}

\begin{figure}[htbp]
\centering
\includegraphics[width=\textwidth]{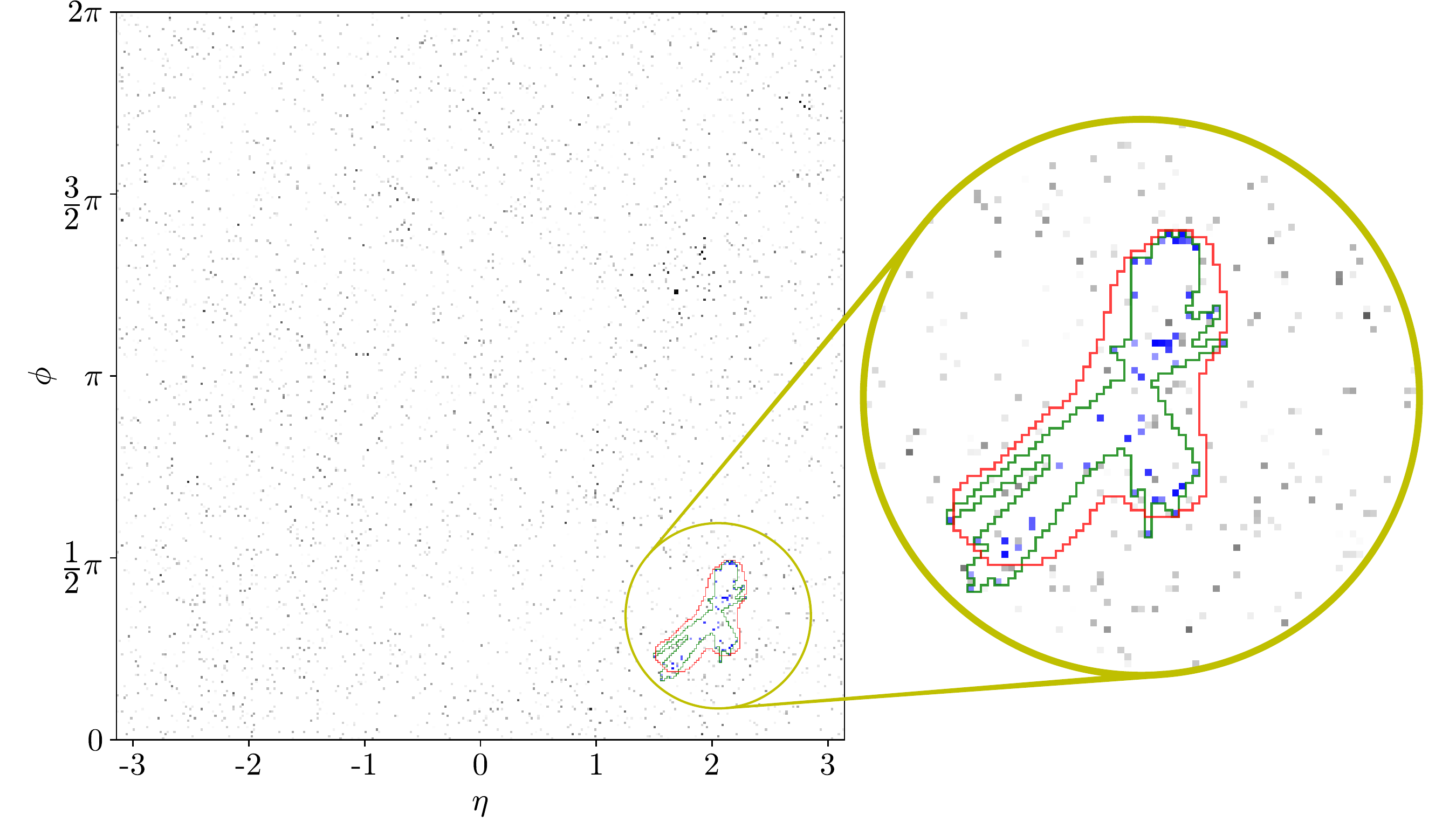}
\caption{A typical event containing one Higgs and three QCD jets (overlaid by pileup events). Each pixel corresponds to $\Delta\eta\times\Delta\phi=0.020\times0.020$. The grayscale of the image are inverted for better perception. Particles of the Higgs jet are highlighted in blue. The yellow circle has $R=0.8$, which is the minimum radius required for a C/A jet to enclose the Higgs decay products completely. This region is magnified three times for clearer visibility. The Higgs boson in this event has $p_T=327$ GeV. This fat jet has $p_T=417$ GeV and $m=209$ GeV prior to any glooming procedure. The green contour indicates the input mask we constructed in radial shape, and the red one indicates the output. Note that we mandatorily enlarge the output by two pixels around its boundary as the final detection result. In this example, particles inside the enlarged contour produce a reconstruction of $m=118$ GeV and $p_T=329$ GeV, as compared to a trimmed jet of $m=133$ GeV and $p_T=336$ GeV.}
\label{fig:test1}
\end{figure}

\begin{figure}[htbp]
\centering
\includegraphics[width=\textwidth]{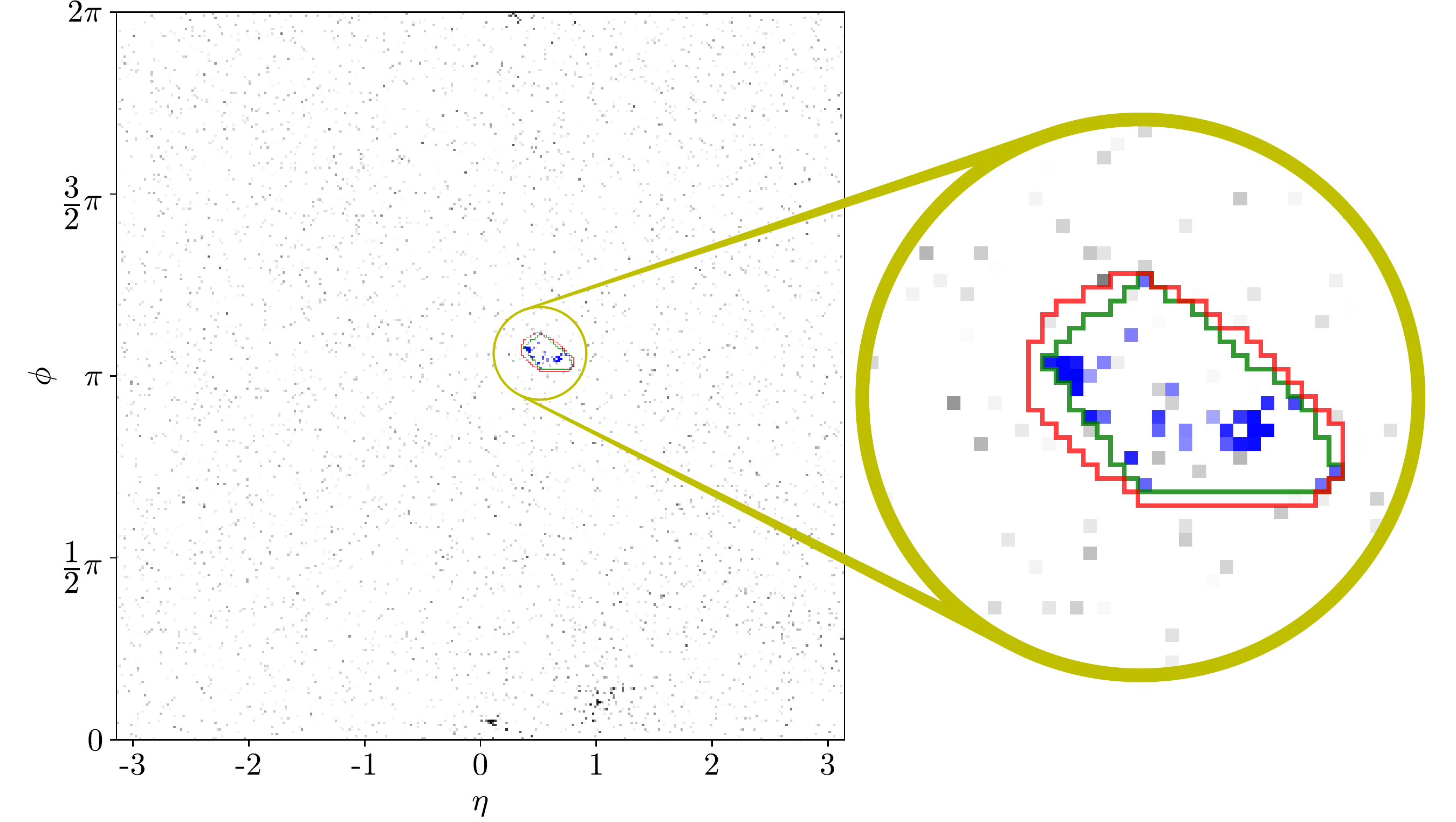}
\caption{Another grayscale-inverted event image. The yellow circle has $R=0.4$, which is magnified six times for clearer visibility. The Higgs boson has $p_T=927$ GeV. The green contour indicates the input mask we constructed in convex hull shape, and the red one indicates the output enlarged by one pixel around its boundary. Reconstructed jets inside the red contour has $m=125$ GeV and $p_T=927$ GeV, as compared to the trimmed one of $m=127$ GeV and $p_T=934$ GeV}
\label{fig:test2}
\end{figure}

Fig.~\ref{fig:test1} and Fig.~\ref{fig:test2} show ground truth labels and test results of two typical events. We compare the performance of our algorithm, denoted as CNN, to a conventional one composed of mass-drop tagger~\cite{Butterworth:2008iy} and trimming~\cite{Krohn:2009th}, denoted as MDT.
In the MDT method, the final state particles are clustered with the Cambridge–Aachen (CA) jet algorithm with appropriate cone size in order to capture most of the Higgs decay products.  They are referred to as fat jets.  
The mass-drop tagger uses two criteria to characterize the substructure of the fat jet: by undoing the jet clustering, there is at least one step which breaks the jet $j$ into two subjets $j_1$ and $j_2$ ($m_{j_1} > m_{j_2}$) such that the mass drop is significant ($m_{j_1} < 0.67 m_{j}$) and the splitting is not asymmetric $\frac{\min (p^2_{T}(j_1), p^2_T (j_2))}{ m^2_j} \Delta R^2_{j_1,j_2} > 0.09$. 
The trimming method selects the hard subjets inside a fat jet, in order to mitigate the pileup contamination. 
We optimize the jet clustering parameter and jet trimming parameters to achieve best reconstruction efficiency within 5 GeV of $m_H$ (same values are obtained for those parameters when optimizing within the 20 GeV mass window). 
The cone size parameters ($R_0$) in jet clustering are 1.8, 1.5, 1.1 and 0.9 for the event samples with $p^H_T$ in the ranges of [200, 250] GeV, [300, 350] GeV, [400, 450] GeV and [500, 550] GeV, respectively. 
The fat jets are trimmed by re-clustering the constituents into $R_\mathrm{sub}=0.20~k_t$-subjets and discarding those with $p_T^\mathrm{subjet}< f_{\rm cut} ~p_T^\mathrm{jet}$, where $f_{\rm cut}=0.05$.
The goal is to reconstruct the four-momentum of a jet as accurately as possible, so we use reconstructed mass, transverse momentum, rapidity and azimuth as the criteria to measure the qualities of jet finding and clustering algorithms. Among them, the distribution of reconstructed mass is the most obvious, as it has ground truth values concentrated at $\lesssim$ 125 GeV shown in Fig.~\ref{fig:cnn_mdt}.

\subsection{Higgs reconstruction efficiency} \label{sec:de}
In detecting the Higgs jets of an event with our method, the pixels under the predicted mask are considered as the constituents of the Higgs jet candidates before pileup subtraction (assuming zero invariant mass for each constituent).  Each connected region of the mask will be regarded as an individual Higgs jet. 
The four-momentum of each Higgs jet candidates is given by the vector sum of four-momenta of its constituents. 

\begin{figure}[htbp]
\begin{center}
\begin{subfigure}[t]{0.37\textwidth}
\includegraphics[width=\textwidth]{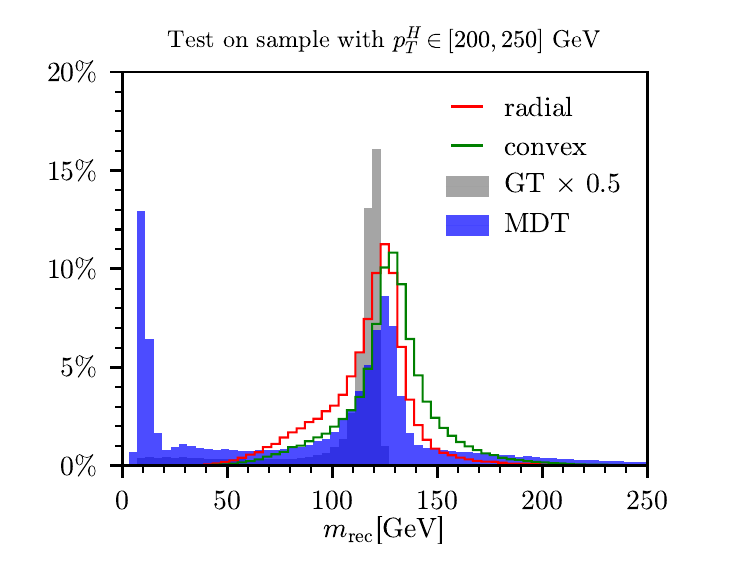}
\end{subfigure}%
\begin{subfigure}[t]{0.37\textwidth}
\includegraphics[width=\textwidth]{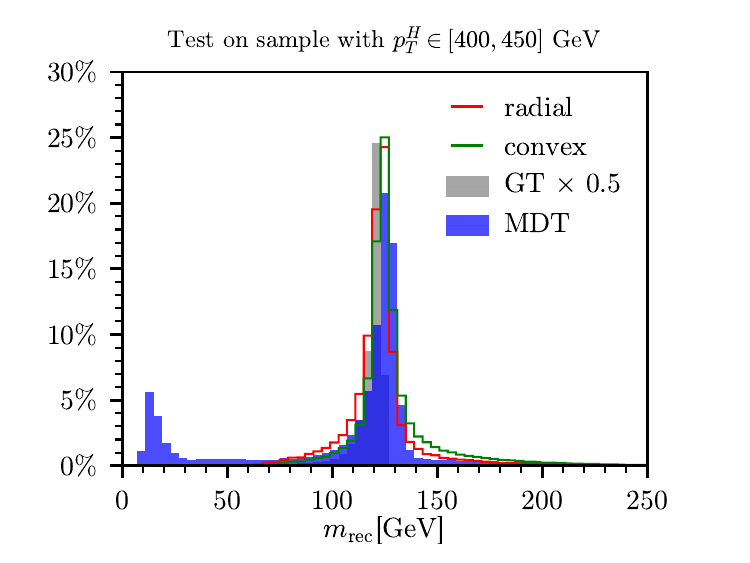}
\end{subfigure}%
\begin{subfigure}[t]{0.37\textwidth}
\includegraphics[width=\textwidth]{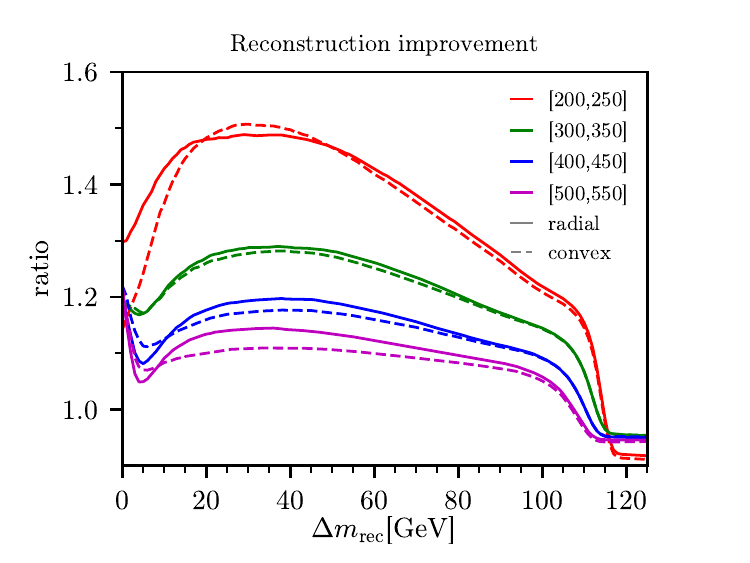}
\end{subfigure}
\end{center}
\caption{Left and middle panels: reconstructed Higgs invariant mass distributions of event samples with $p_T^H$ in the ranges of [200, 250] GeV and [400, 450] GeV. Right panel: reconstruction efficiency improvement of CNN over MDT, in terms of the number of reconstructed Higgs within $\Delta m$ of $m_H$. 
The heights for the distributions of ground truth Higgs mass are rescaled by a factor of 0.5 for visibility.}
\label{fig:cnn_mdt}
\end{figure}

Fig. \ref{fig:cnn_mdt} shows the invariant mass distributions of reconstructed Higgs bosons through CNN and MDT methods.
Both of CNN and MDT may find multiple Higgs jet candidates in one event, in that case, we keep the one with the highest score for CNN and take the one closest to the true Higgs boson for MDT. In practice, this can not be realized. But one can apply more sophisticated Higgs tagging method to identify the correct Higgs jet. We take the ideal case for comparison purpose.
Applications to the event samples with $p_T^H$ in the ranges of [200, 250] GeV and [400, 450] GeV are illustrated. 
We conclude from the plots that: 1) Our methods with both radial mask and convex mask outperform the MDT method; (2) The CNN trained using convex mask tends to predict higher Higgs invariant mass than the CNN trained using radial mask, due to the larger jet area, i.e. higher pileup contamination; (3) The improvements of CNNs over MDT are more significant for Higgs jets with lower $p_T$; (4) all methods (MDT and CNNs with radial and convex mask) predict higher Higgs mass than the ground truth Higgs jet. 
The right panel of Fig. \ref{fig:cnn_mdt} shows the efficiency boost, defined as 
\begin{align}
\text{ratio} = \frac{\text{Number of events with } m_\text{rec}\in [m_H-\Delta m, m_H+\Delta m] \text{ in CNN method}}{\text{Number of events with } m_\text{rec}\in [m_H-\Delta m, m_H+\Delta m] \text{ in MDT method}}
\end{align}
The CNNs with two definitions of mask achieve similar amount of improvement and the improvements are more significant for lower $p_T$ samples. However, due to the pileup effects, the peaks are shifted toward higher values, leading to the drop of ratio at $\Delta m \sim 5$ GeV for events samples with $p_T \gtrsim 300$ GeV.
If we keep events with $m_\text{rec}\in [105, 145]$ GeV, CNN methods yield 10\%-50\% more signal events than the MDT method. 

\begin{figure}[htbp]
\centering
\includegraphics[width=0.32\textwidth]{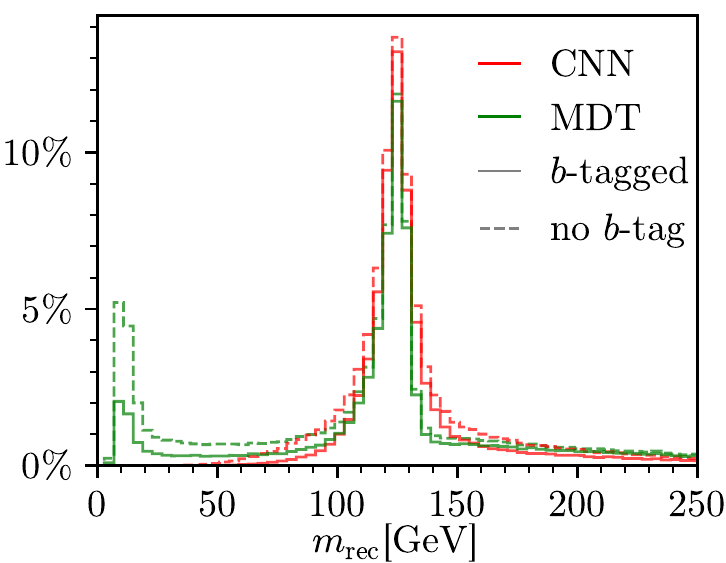}
\includegraphics[width=0.32\textwidth]{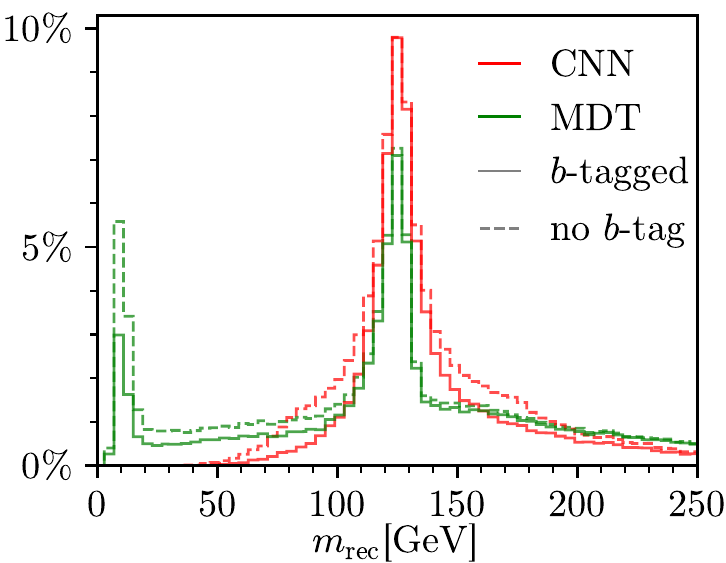}
\includegraphics[width=0.32\textwidth]{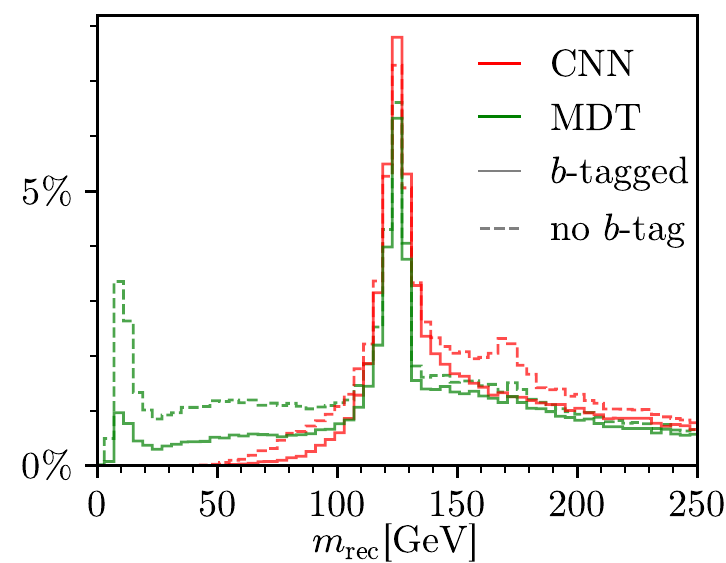}
\caption{The reconstructed Higgs invariant mass distributions from CNN and MDT methods being tested on the event samples of the $HHjjj$ (left panel), $Ht\bar t$ (middle panel) and hypothetical SUSY process (right panel). Solid lines represent results with the $b$-tagger being applied to the Higgs jet candidates. }
\label{fig:others}
\end{figure}

Although the network is trained on a single Higgs process, i.e., one Higgs boson plus three QCD jets, it may serve as a general Higgs tagger in all kinds of events. For demonstration purposes, we showcase its capability in three other processes which contain at least one Higgs boson: 1) two Higgs boson plus three QCD jets; 2) one Higgs boson plus two top quarks; 3) a hypothetical SUSY process, $pp\to\tilde t^*_1\tilde t_1\to\bar t\tilde \chi_1^0 t \tilde \chi_2^0\to \bar t\tilde \chi_1^0 t H \tilde \chi_1^0$, where $m_{\tilde \chi_1^0}=100$ GeV, $m_{\tilde \chi_2^0}=800$ GeV and $m_{\tilde t_1}=1$ TeV. The Higgs bosons in all processes are forced to be boosted, $p_T>200$ GeV and decay to two $b$-quarks. We apply the network trained with radial mask directly to these scenarios and find promising performance of the Higgs detection. 
The normalized Higgs mass distributions are shown in Fig.~\ref{fig:others}. 
As a comparison, distributions of trimmed jets (the one closest to the true Higgs boson in the MDT analysis) are also plotted. 
Note for the $HH$+jets process, two Higgs jet candidates are used in CNN (two highest score jets) and MDT (each is closest to one of the true Higgs boson) analyses. 
The parameters for jet clustering and trimming are shown in Table.~\ref{tab:trim}. Again we optimize the MDT parameters to achieve the best reconstruction efficiency within 5 GeV of $m_H$.
We also adopt the b-tagging requirement for the Higgs jet candidate, which is found to be helpful in suppressing fake Higgs jet from top quark for the CNN method.  
A subjet is considered to be $b$-tagged if there are any bottom quarks with $p_T>20$ GeV lie within a cone of radius $R=0.2$ around the subjet direction. 
Solid lines in the plots represent doubly $b$-tagged jets (two subjets being b-tagged). 
Here for simplicity, we assume a 100\% efficiency of $b$-tag.

\begin{table}[t]
\centering
\begin{tabular}{cccc}
\hline
process&$R_0$&$R_\text{sub}$&$f_\text{cut}$\\
\hline
$HHjjj$&1.4&0.19&0.05\\
$Htt$&1.4&0.19&0.05\\
$\tilde t^*_1\tilde t_1$&0.9&0.18&0.05\\
\hline
\end{tabular}
\caption{Jet clustering and trimming parameters in the MDT method, as introduced at the beginning of Sec.~\ref{perform}.  \label{tab:trim}}
\end{table}

\begin{figure}[htbp]
\centering
\includegraphics[width=0.45\textwidth]{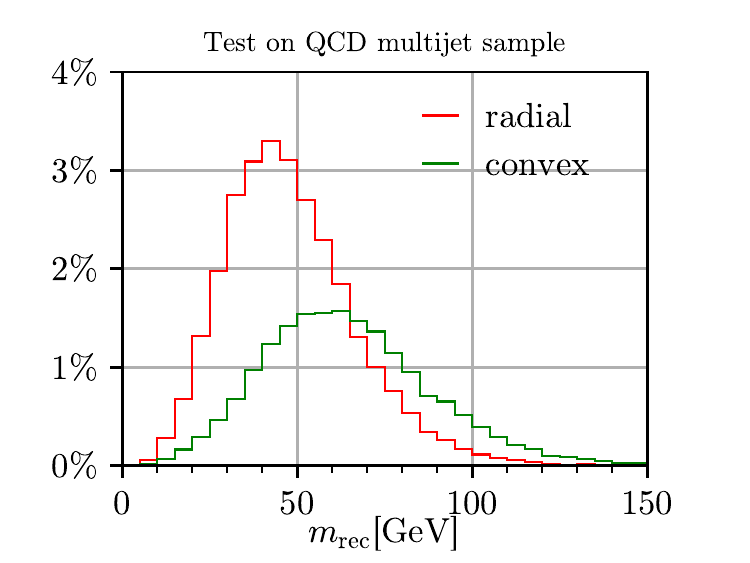}
\includegraphics[width=0.45\textwidth]{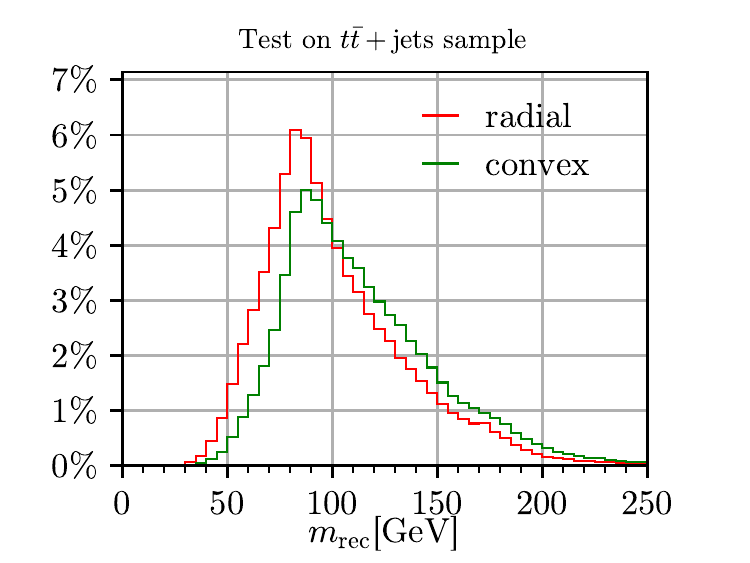}
\caption{The reconstructed (highest score) Higgs invariant mass distributions from our networks being tested on the QCD multijet sample and $t\bar{t}+$jets sample. Only events which contain at least one CNN tagged Higgs jet with score greater than 0.9 are used. \label{cnn:qcd}}
\end{figure}

Finally, we apply our network to non-Higgs processes to investigate whether the reconstructed Higgs invariant masses of backgrounds are also shaped. 
For illustration, the QCD multijet process (matched to parton shower up to three jets with $p_T> 20$ GeV) and the $t\bar{t}$ process (matched to parton shower up to two jets) are considered. 
Taking events which contain at least one CNN tagged Higgs jet with score greater than 0.9, the invariant mass distributions of the highest score Higgs candidate for those two processes are shown in Fig.~\ref{cnn:qcd}. 
The selection ($\text{score}>0.9$) efficiencies of the CNN with convex mask and radial mask are 18.3\% (69.3\%) and 28.1\% (74.7\%) for the QCD multijet ($t\bar{t}$) process, respectively. 
The distributions of the fake Higgs invariant mass are peaked at around 50 GeV and 80 GeV for the QCD and $t\bar{t}$ backgrounds, reflecting the typical jet energy in those two processes before pileup subtraction. 
More details about background rejection will be discussed in Sec.~\ref{sec:sigbkg}. 

\subsection{Higgs reconstruction accuracy} \label{sec:recopile}

\begin{figure}[t]
\centering
\begin{subfigure}[t]{0.48\textwidth}
\includegraphics[width=\textwidth]{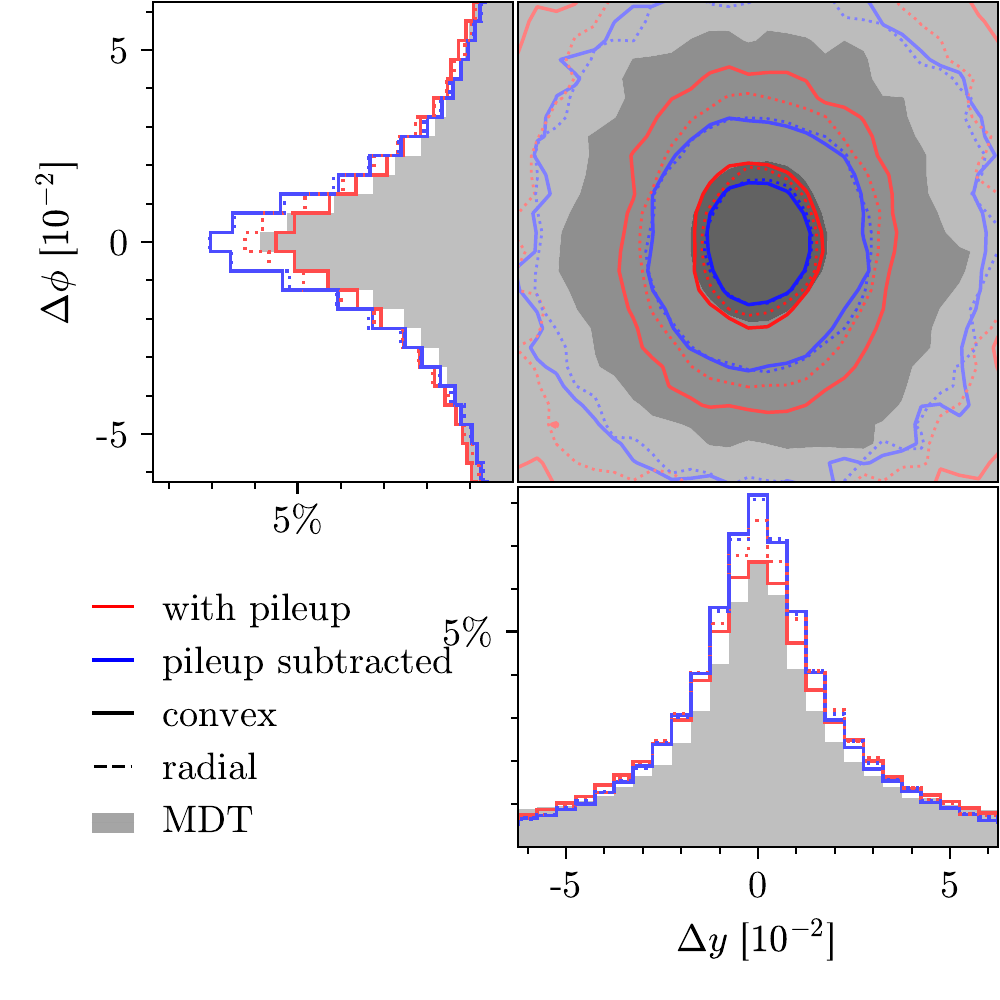}
\end{subfigure}
\enskip
\begin{subfigure}[t]{0.48\textwidth}
\includegraphics[width=\textwidth]{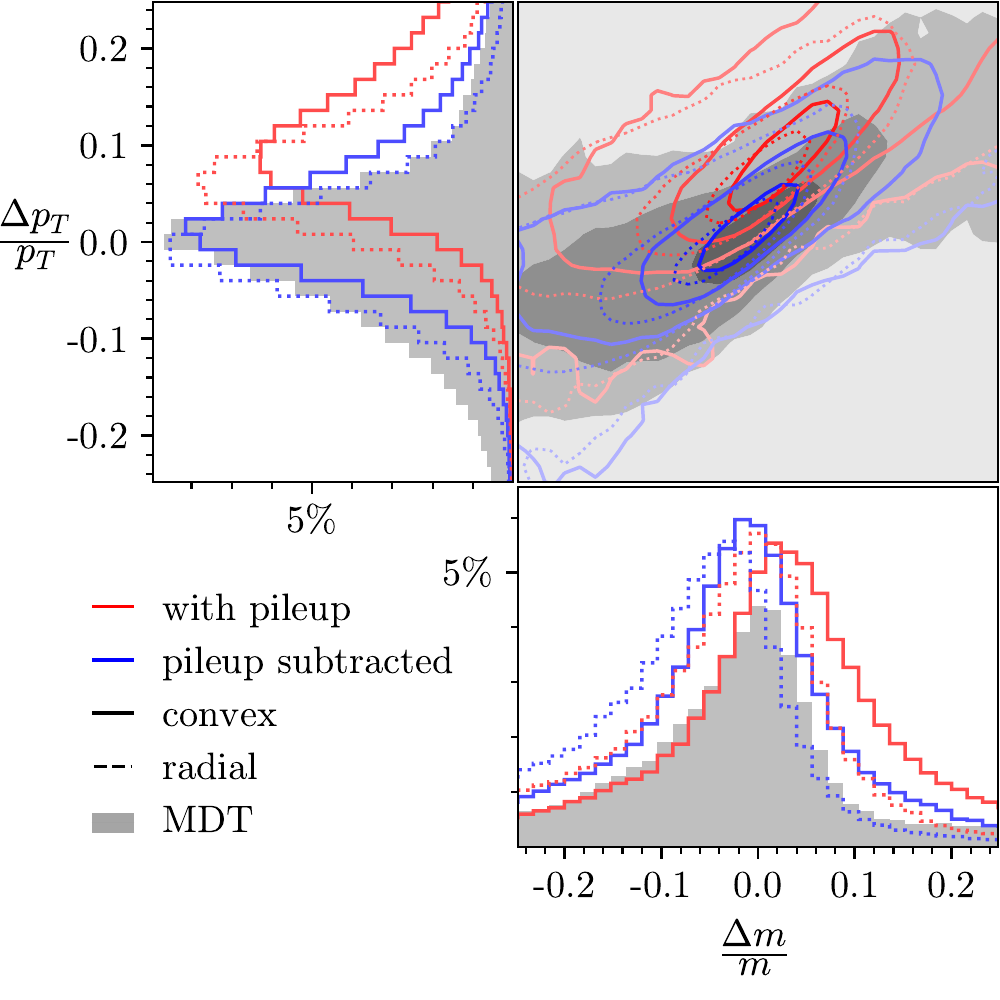}
\end{subfigure}
\par\smallskip
\begin{subfigure}[t]{0.48\textwidth}
\includegraphics[width=\textwidth]{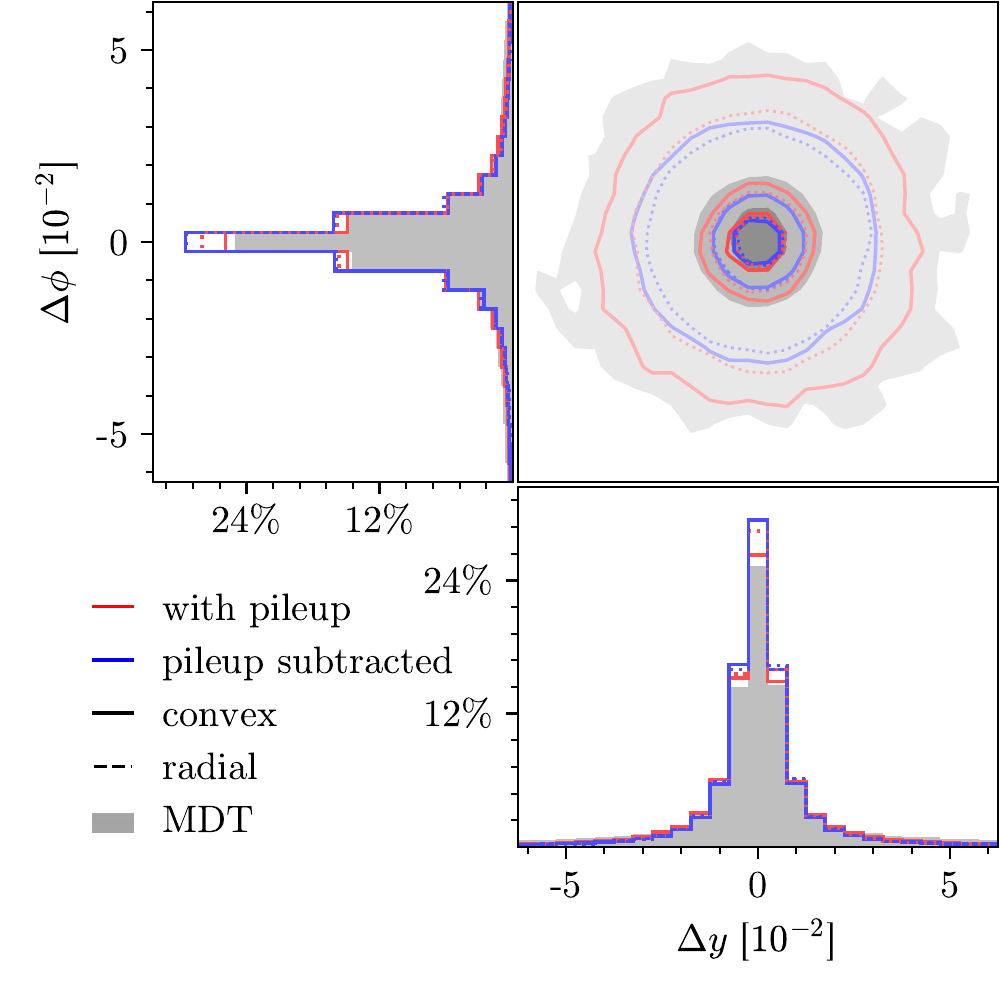}
\end{subfigure}
\enskip
\begin{subfigure}[t]{0.48\textwidth}
\includegraphics[width=\textwidth]{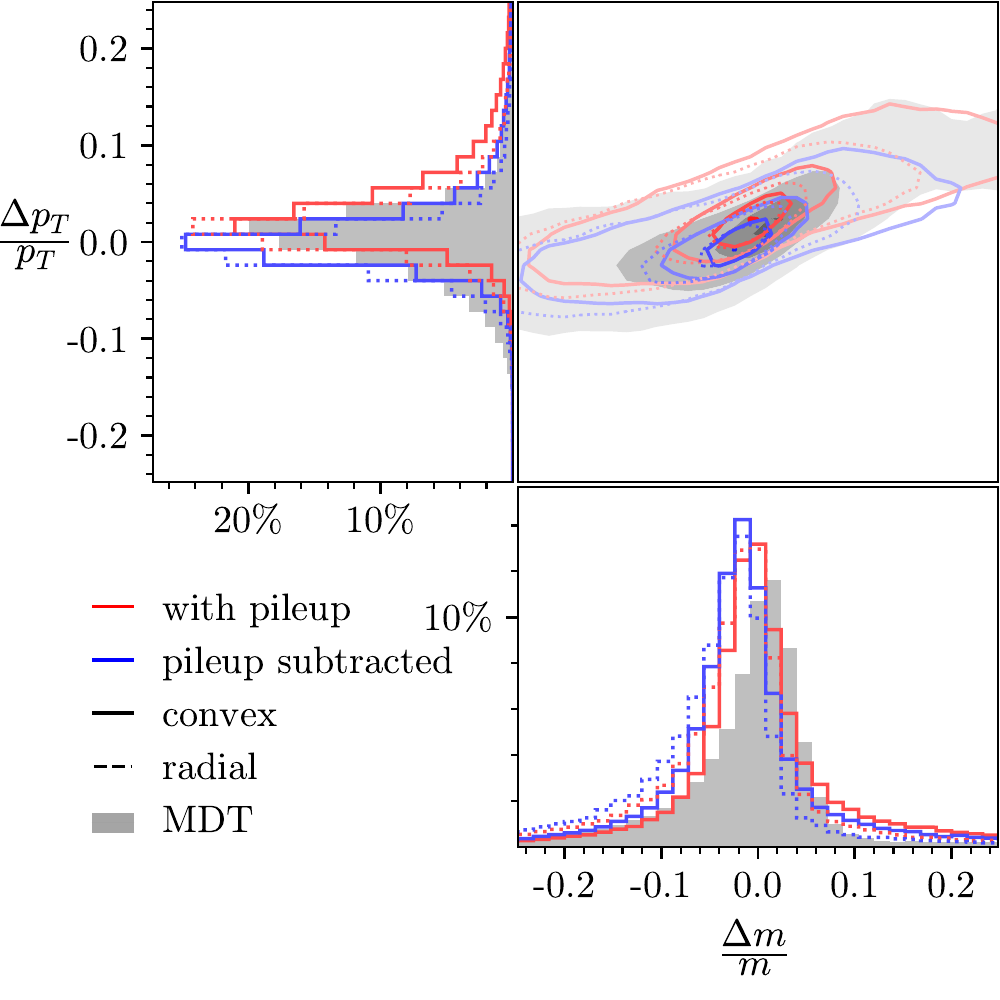}
\end{subfigure}
\caption{Higgs reconstruction accuracies of the MDT method (gray regions), and CNN method with convex (red solid contours) and radial (red dashed contours) mask. The performances of CNN after pileup subtraction are shown by blue contours. Upper panels: event sample with Higgs transverse momentum in the range [200, 250] GeV. Lower panels: event sample with Higgs transverse momentum in the range [400, 450] GeV
}
\label{fig:err_rad}
\end{figure}

To show that detection and segmentation produce better accuracy than clustering, tagging and trimming, we present the deviations of reconstructed variables (including rapidity $y$, azimuth $\phi$, invariant mass $m$ and transverse momentum $p_T$) in Fig.~\ref{fig:err_rad}, where the distributions are normalized such that the sum of all simulated events for each case is equal to one.
Two event samples with Higgs boson $p_T$ in the ranges of [200, 250] GeV and [400,450] GeV are tested for illustration. 
Only correctly tagged Higgs bosons are taken into account. We consider it a correct tag if the true Higgs boson falls within $R_0$ (for MDT) of the reconstructed one or falls within the mask (for CNN). Note that these criteria can not be applied to actual event selections, hence they are only suitable for demonstration of reconstruction accuracy. Different shades of gray regions and colored contours indicate 20\%, 40\%, 60\% and 80\% of events, respectively. The closer they are to the center, the higher accuracy they stand for.

We find pileup mitigation is essential in reconstructing the $p_T$ of Higgs jet precisely with the CNN method.  
The pileup contamination in the masks can be subtracted by using the $p_T$ density based method~\cite{Cacciari:2007fd}. 
For as many as 50 pileup events, their $p_T$ distribute almost uniformly on the $\eta \times \phi$ plane. Let the density of $p_T$ be $\rho$. According to the mask shape of each jet, we subtract pileup from the original jet, then obtain the corrected four-momentum. The corrections of mass and $p_T$ are roughly,
\begin{equation}
\delta p_T\simeq -\rho A
\end{equation}
\begin{equation}
\delta m^2\simeq -\rho p_T A\langle \Delta R^2 \rangle\gtrsim -\rho p_T A\dfrac{\Delta R^2_\text{max}}{2}
\end{equation}
where $A$ is the area of the mask, $\Delta R$ represent the distance between a point on the mask and the Higgs boson, and $\langle \Delta R^2\rangle$ is the area-averaged $\Delta R^2$. Note that $\Delta R_\text{max}$ does not equal half of the diameter. 
According to our simulation, the $p_T$ density $\rho\simeq 55-60$ GeV. 
However, we find setting $\rho=35$ GeV and imposing $m=0$ GeV on subtracted four-momentum of each pixel provide the lowest deviation of the reconstructed Higgs $p_T$, as shown in Fig.~\ref{fig:err_rad}. 
The reasons for choosing smaller $\rho$ are threefold: (1) the masks predicted by CNN do not cover all jet constituents; (2) the mass and $p_T$ of pileup contributions are overestimated when substituting a continuous distribution for a discrete one; (3) the assumption of uniformly distributed $p_T$ from pileup contribution may not valid for some individual events. 


In Fig.~\ref{fig:err_rad}, the grey regions indicate the distributions of reconstructed variables from the MDT method (with trimming imposed). And 
the red solid (dashed) contours correspond to the distributions from the CNN method with convex (radial) mask before pileup mitigation.  
Two CNN methods provide slightly higher precision in rapidity and azimuth reconstruction than the MDT method: 
about 40\% of $H+$jets events can be reconstructed with $|\Delta \phi| \sim |\Delta \eta| < 4 \times 10 ^{-2}$ and  $2 \times 10 ^{-2}$ for the event samples with $p_T^H \in [200, 250]$ GeV and $p_T^H \in [400, 450]$ GeV.
The performances of CNNs are worse than that of the MDT method in terms of $p_T$ reconstruction for lower $p_T^H$ sample, where a systematic excess is prominent. Because the bare jets reconstructed from the CNNs do not include pileup mitigation, other than trying to suppress its contamination via small area of jet mask. 
On average, CNNs reconstruct Higgs mass more accurately than MDT (higher fraction of events around $\frac{\Delta p_T}{p_T} \sim 0$ and $\frac{\Delta m}{m} \sim 0$), even though the deviation of invariant mass is much larger than that of $p_T$. 
The blue contours show the deviations of reconstructed momentum variables after applying the pileup subtraction. 
We can find the accuracy of angular reconstruction can be improved even further. 
About 40\% of $H+$jets events can be reconstructed with $|\Delta \phi| \sim |\Delta \eta| < 3 \times 10 ^{-2}$ and  $1 \times 10 ^{-2}$ for the event samples with $p_T^H \in [200, 250]$ GeV and $p_T^H \in [400, 450]$ GeV, respectively.
The accuracy of $p_T$ reconstruction is greatly improved for both event samples, since it is our criteria for choosing the $\rho$ parameter. 
However, with the simple $p_T$ density based subtraction method, the invariant masses reconstructed by the CNN methods are underestimated. 
The problem is more severe for radial mask and lower $p_T$ Higgs. 

\begin{figure}[htbp]
\centering
\includegraphics[width=0.45\textwidth]{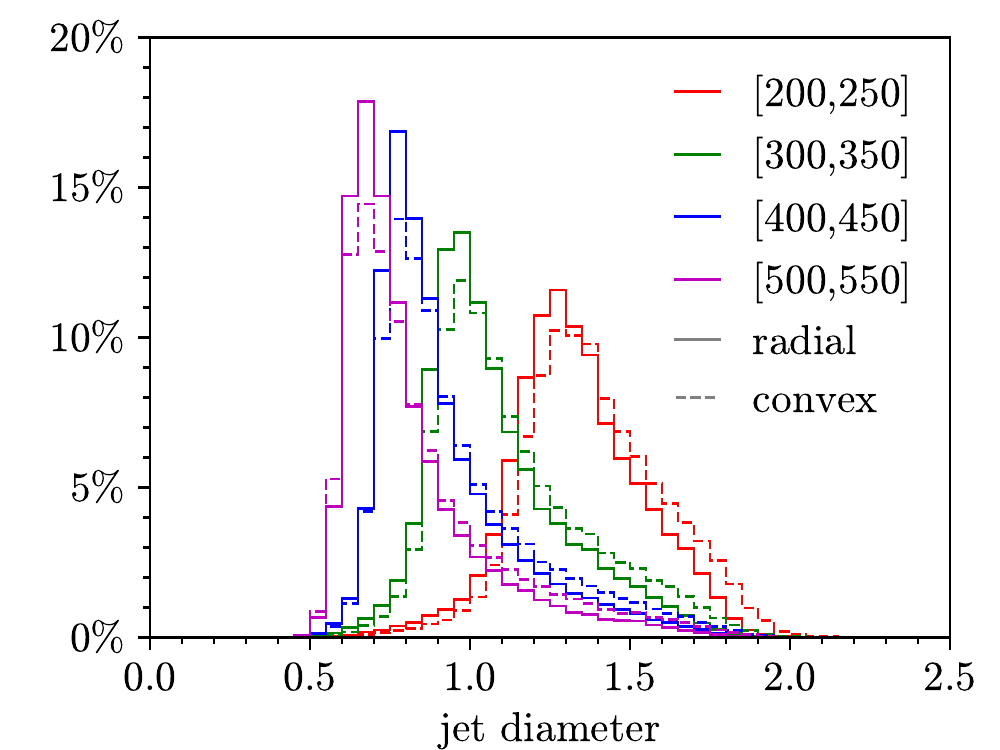}
\includegraphics[width=0.45\textwidth]{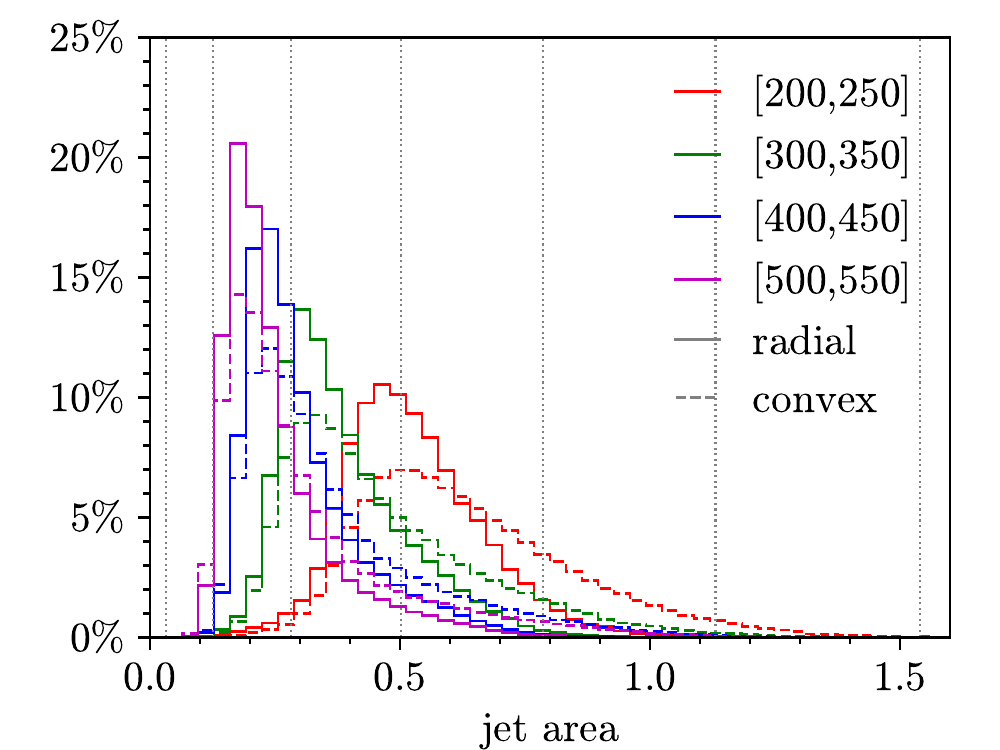}
\caption{Left panel: jet diameter distribution. Right panel: jet area distribution, vertical dotted lines indicate circular areas with $R=0.1,0.2,\ldots,0.7$ (from left to right). Colors indicate the range of $p_T^H$ of the event sample. Solid (dashed) lines correspond to the case of radial (convex) mask.}
\label{jet_area}
\end{figure}

The CNN with radial mask achieves higher accuracy than the CNN with convex mask for all reconstructed variables, since it represents more precise shape of the Higgs jet. 
The diameters (defined as the largest angular distance $\Delta R=\sqrt{(\Delta \eta)^2+(\Delta \phi)^2}$ between any two marked particles) and areas of Higgs jets predicted by CNN are shown in Fig. \ref{jet_area}. 
The convex mask has slightly longer diameter and larger jet area than the radial mask.
The peak of jet diameter distribution is roughly given by $(2.1\sim 2.8) \times m_H/ p_T^H$. 
In the right panel of Fig.~\ref{jet_area}, the vertical lines indicate circular areas with $R=0.1,0.2,\ldots,0.7$ (from left to right), corresponding to the jet areas obtained by the anti-$k_T$ algorithm with cone size parameter $R$.  
For Higgs jet with transverse momentum $p_T^H>200$ GeV, the area of masks are $\lesssim 1.0$, corresponding to the anti-$k_T$ jet with cone size $R \sim 0.5-0.6$, which is much smaller than the one used in MDT (to capture the Higgs jet with $p_T^H\sim 200$, $R=1.0-1.2$ should be chosen). The feature is useful to suppress the pileup contamination. 

\subsection{Signal and background discrimination} \label{sec:sigbkg}

At the end of Sec.~\ref{sec:de}, we have shown the distributions of the fake Higgs invariant mass obtained from applying the CNN on the non-Higgs processes, including QCD multijet and $t\bar{t}$.  
In this section, we 
examine the discrimination power of our trained networks on a dataset composed of two top quarks with up to two QCD jets events. Ideally, the score returned by the network alone should suffice as a discriminant between real and fake Higgs. With a varying number between 0 and 1 as the threshold, one would get a receiver operating characteristic (ROC) curve displaying how well the signals and backgrounds can be differentiated. Unfortunately, our network is not powerful enough yet to accomplish this mission in one step, we will probably get further improvements with a more sophisticated representation of the event, network with more branches and also need to train the network with the non-Higgs jets (such as top quark jet in this case) and their masks, to get further improvements. 
For now, we complement the outputs of our network with a few substructure variables and implement a Boosted Decision Tree (BDT)~\cite{Roe:2004na,Hocker:2007ht} to fulfill this task. Variables in consideration are listed in Table. \ref{tab:bdt}. 
In addition to the transverse momentum ($p_T$), invariant mass $(m)$ and their ratio ($p_T /m$), the following variables are adopted in the BDT analyses~\cite{Aad:2014rga}:
\begin{itemize}
\item $p_{T,\max}$: the transverse momentum of the hardest constituent inside a cone of
radius $R=0.2$ centered on the direction of the Higgs jet candidate; 
\item $f_{\text{center}}=E^{(0.1)}/E^{(0.2)}$ where $E^{(0.1)}$ ($E^{(0.2)}$) is the
total energy of constituents in a cone of radius $R=0.1$ (0.2) centered on the direction of the Higgs jet candidate;
\item $R_{\rm weight}$: the $p_T$-weighted sum of the angular distances of all constituents inside the Higgs jet candidate;
\item the $N$-subjettiness variable
$\tau_{21} = \tau_2/\tau_1$ and $\tau_{32}=\tau_3/\tau_2$ that allow one to resolve the substructure of
the Higgs jet candidate~\cite{Kim:2010uj,Thaler:2010tr}, with $\tau_N$ being defined by
\begin{equation}
  \tau_N =\frac{\sum_k \min \{ \Delta R_{1,k}, \Delta R_{2,k}, \dots, \Delta R_{N,k} \}}{\sum_k p_{T,k} R_0}.
\end{equation}
In this expression, the summations run over all of the Higgs jet constituents, $R_0$ is the jet cone size parameter in the fat jet clustering algorithm and $\Delta R_{I,k}$ denotes the distance in the transverse plane between the subjet candidate $I$ and the jet constituent $k$.
\end{itemize} 

The BDT method uses 500 tree ensemble that requires a minimum percentage of training events in each leaf node of 2.5\% and a maximum cell tree depth of three. The rest of the parameters are set to default ones in the TMVA package~\cite{Hocker:2007ht}. 
It is trained on the 80\% of the reconstructed Higgs jet candidate from $Hjjj$ (with $p_T(H)>200$ GeV) and $t\bar{t}$ events and is test on the rest of them. The Kolmogorov-Smirnov test~\cite{Chakravarty:109749} of training sample and testing sample should be greater than 0.01 to avoid overtraining.

\begin{figure}[htbp]
\centering
\includegraphics[width=0.4\textwidth]{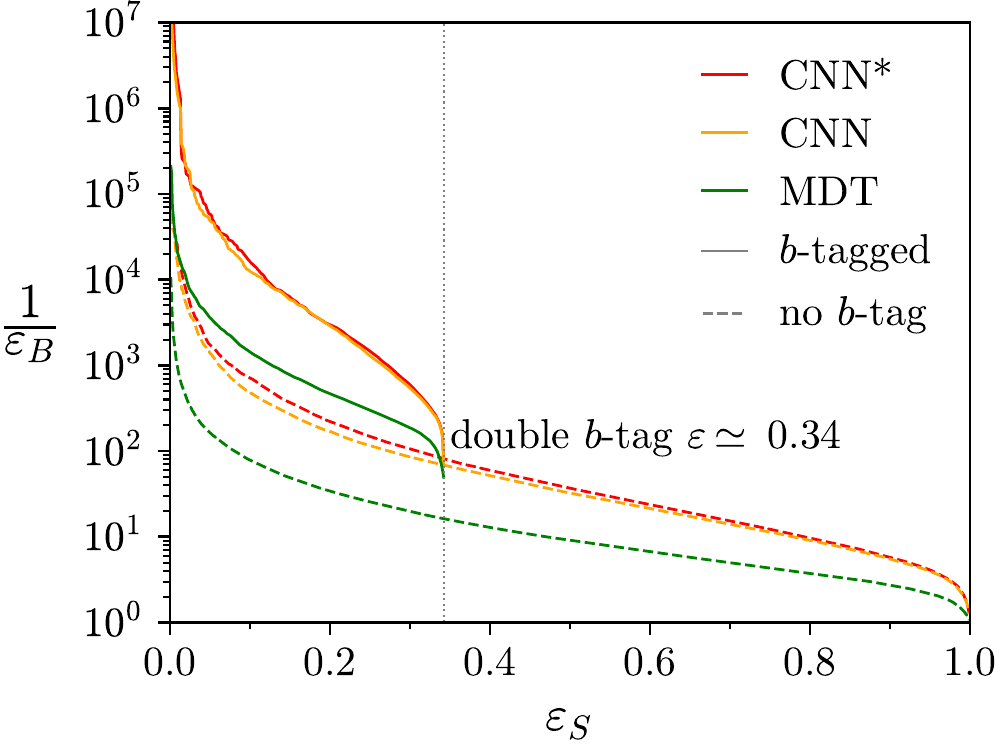}
\enskip
\includegraphics[width=0.4\textwidth]{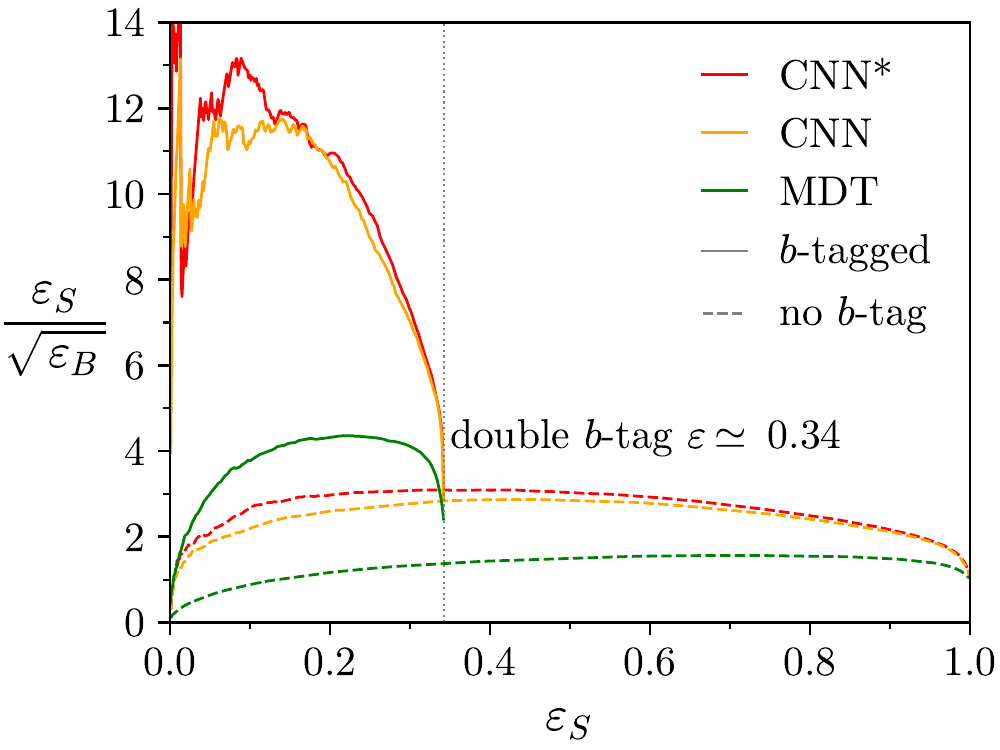}
\caption{Left panel: receiver operating characteristic (ROC) curves. Right panel: significance improvement characteristic (SIC) curves.}
\label{fig:roc}
\end{figure}

As before, we conduct a comparison study on the performances of the MDT and CNN method. The performance of the MDT (CNN) method correspond to the performance of the BDT that uses variables checked as in the second (third) column of Tab.~\ref{tab:bdt}. Moreover, we have conducted a BDT analysis dubbed CNN$^*$, which includes diameter and area of the reconstructed Higgs jet, in addition to those variables in the CNN analysis. The leading $p_T$ fat jet with substructure and the masked jet with highest score are considered as the Higgs jet candidates in the MDT and CNN method, respectively. 
The ROC curves and significance improvement curves (SICs) are shown in Fig. \ref{fig:roc}, where solid lines indicate doubly $b$-tagged events. 
Here, we consider it a $b$-tag if the jet or subjet with a certain radius cover the $\eta-\phi$ coordinates of a $b$ quark with $p_T>20$ GeV. On top of that, a $b$-tag efficiency of 70\%, $c$-quark mis-tagging rate of 15\% and the other light quark and gluon mis-tagging rates of 0.8\% are set. The overall efficiency of double $b$-tag is about 34\%, so the ROC curve does not exceed $\varepsilon\simeq0.34$. 
From the ROC, we find that the background rejection for a given signal selection efficiency can be improved by an order of magnitude once the Mask R-CNN score is included in the BDT analysis. 
Equipped with the Mask R-CNN score and b-tagging, the signal significance factor can achieve $\sim \mathcal{O}(10)$ with signal efficiency $\epsilon_S \lesssim 0.25$.  
Note that the SIC is an approximate proxy to the significance in an actual analysis. 
It will show instabilities when the event samples are inadequate~\cite{Gallicchio:2010dq}.

\begin{table}
\centering

\begin{tabular}{cccc}
\hline
variable&MDT&CNN&CNN*\\
\hline
$p_T$&\checkmark&\checkmark&\checkmark\\
$m$&\checkmark&\checkmark&\checkmark\\
$p_T/m$&\checkmark&\checkmark&\checkmark\\
$p_{T,\text{max}}$&\checkmark&\checkmark&\checkmark\\
$f_\text{center}$&\checkmark&\checkmark&\checkmark\\
$R_\text{weight}$&\checkmark&\checkmark&\checkmark\\
$\tau_2/\tau_1$&\checkmark&\checkmark&\checkmark\\
$\tau_3/\tau_2$&\checkmark&\checkmark&\checkmark\\
score&&\checkmark&\checkmark\\
diameter&&&\checkmark\\
area&&&\checkmark\\
\hline
\end{tabular}
\caption{Variables that are adopted in the BDT analyses. The definitions for those variables can be found in the text. \label{tab:bdt}}
\end{table}

\section{Conclusion and outlook} \label{sec:conclusion}

The Mask R-CNN framework is adopted to reconstruct the Higgs jets from the Higgs production events, each is overlaid with 50 pileup events. 
The method is capable of reconstructing Higgs jets in a wide range of $p^H_T$, mainly owing to the component of FPN. 

The Mask R-CNN is trained on event samples of the $H+$jets process with flat Higgs transverse momentum ($p_T^H$) distribution in the range of [200, 600] GeV. 
The Higgs jet in an event is defined by a mask built according to the truth level information obtained from simulation. Two schemes of mask definition are proposed in this work: convex mask and radial mask. 
Benefited from more accurate jet shape, the CNNs with both definitions of mask can reconstruct the Higgs angular variables more precisely than the MDT method as shown in the left panels of Fig.~\ref{fig:err_rad}. 
About 5\%-10\% more events with $\Delta \eta$ and $\Delta \phi$ in [-0.05,0.05] are regained in the CNN method.
The deviation of the reconstructed Higgs $p_T$ can be reduced substantially after applying the pileup subtraction. For the sample with $p^H_T \in[200,250]$ ([400,450]) GeV, the peak of deviation reduce from $\Delta p_T/p_T \sim 0.08~(0.02)$ to $\Delta p_T/p_T \sim 0 ~(0)$. 
On the other hand, the accuracy of the reconstructed Higgs invariant mass is promising before the pileup subtraction. The over-simplified $p_T$ density based pileup subtraction renders the reconstructed Higgs invariant mass shifting toward lower values (with peak at $\Delta m/m \sim -0.02$ for the sample with $p^H_T \in[200,250]$). 
The deviation of the invariant mass is less severe when the Higgs transverse momentum is higher . 

Even though the Mask R-CNN is trained on the events of the $H$+jets process, it is totally applicable to other processes. For illustration, three different processes are considered in the work: 1) two Higgs bosons plus three QCD jets; 2) one Higgs boson plus two top quarks; 3) a hypothetical SUSY process, $pp\to\tilde t^*_1\tilde t_1\to\bar t\tilde \chi_1^0 t \tilde \chi_2^0\to \bar t\tilde \chi_1^0 t H \tilde \chi_1^0$. Promising Higgs detection efficiencies are obtained for all processes. 
In particular, it is found that the Mask R-CNN training on single Higgs events is capable of reconstructing both Higgs jets in $HH+$jets sample. 
We also demonstrate the signal and background discrimination capacity of our network by applying it to the $t\bar{t}$+jets process, which is usually the dominant background in Higgs-related searches. 
Taking the outputs of the network as new features to complement traditional jet substructure variables, the background rejection rate can be increased by an order of magnitude while keeping the same signal acceptance rate.

The method proposed in this work can be generalized to detect multiple different objects (such as W/Z boson jets) simultaneously. Moreover, the $\eta$, $\phi$ and $p_T$ of the Higgs jet are obtained by the vector sum of momenta of all marked particles (pixels). One could integrate the momentum regression into the network. It should be noted that two schemes of mask definition given in this work are not unique and may not the best. They may be improved in realistic experiments.
As for jet induced by a colored particle, due to its color connection with other partons and with the beam remnant, the assignments of the final state particles are ambiguous. 
It will be more difficult to define an appropriate mask for them. 
Training our network also with masks of top quark jets could further enhance the signal and $t\bar{t}$+jets background discrimination power. 
We leave those points to future works.

~\\
\textbf{Note Added}

While we were preparing this manuscript, a paper with similar purpose~\cite{Ju:2020tbo} appears on the arXiv. Different neural network was adopted in their study. The effects from pileup events as well as the applications to processes other than the one used for training are not studied in their work.

\begin{acknowledgments}
We thank Jun Guo for valuable comments and suggestions. 
This work was supported in part by the Fundamental Research Funds for the Central Universities, by the National Science Foundation of China under Grant No. 11905149, by Projects No. 11847612 and No. 11875062 supported by the National Natural Science Foundation of China, and by the Key Research Program of Frontier Science, Chinese Academy of Sciences.
\end{acknowledgments}

\bibliographystyle{jhep}
\bibliography{ref}
\end{document}